\documentclass[12pt]{article}

\usepackage{epigraph}
\usepackage{appendix}
\usepackage{float}
\usepackage{amssymb}

\input{A_PreambleSettings.tex}

\title{\textbf{ \Large 
Peer Influence on West Point Cadets' Civil War Allegiances}\thanks{YG, MJ, and RJ designed the research.  YG collected the data.   YG, MJ, and RJ analyzed the data.   YG, MJ, and RJ wrote the paper.  
We thank Ran Abramitzky, Hoyt Bleakley, Jim Fearon, Sebastián Villamizar Santamaría, Edward Glaeser, Joel Mokyr, Sandip Sukhtankar,  Gavin Wright, the editor, and the reviewers, for helpful comments. We are grateful to Junsheng Huang for excellent research assistance.}}
\author{Yuchen Guo\thanks{Corresponding Author, Chinese University of Hong Kong; guoych53@gmail.com.} \and Matthew O. Jackson$^*$\thanks{Corresponding Author, Stanford University and the Santa Fe Institute; jacksonm@stanford.edu.} \and Ruixue Jia\thanks{Corresponding Author, UCSD, CEPR and NBER; rxjia@ucsd.edu.} }
\date{\today}

\begin{document}
	\setlength{\parskip}{0.05in}
	\setlength{\parindent}{1.5em}

	\maketitle

\begin{abstract}
\noindent 
Do social networks and peer influence shape major life decisions in highly polarized settings? We explore this question by examining how peers influenced the allegiances of West Point cadets during the American Civil War. 
Leveraging quasi-random variations in the proportion of cadets from Free States, we analyze how cadets' decisions about which army to join depended on the composition of their peers. We have three main findings.  
First, there was a strong and significant peer effect: a higher proportion of classmates from Free States significantly increased the likelihood that cadets from Slave States joined the Union Army. Second, the peer effect varies with geography, most notably with the slave population share in cadets' home states or counties, and with cadets' own slave ownership in 1860. Third, peer effects were amplified by shared experiences such as having served together in the Mexican-American War, continuous military service, and belonging to the same cohort, suggesting that sustained interaction is important.

\end{abstract}

\clearpage

\epigraph{
``... for Civil War soldiers, the group cohesion and peer pressure that were powerful factors in combat motivation were not unrelated to the complex mixture of patriotism, ideology, concept of duty, honor and manhood, and community or peer pressure that prompted them to enlist in the first place.''
}{James M. McPherson, \textit{For cause and comrades: Why men fought in the Civil War \citeyearpar{mcpherson1997cause}}}

\section{Introduction}

Do social networks and peer influence shape individuals' choices when societies are polarized? This question is central not only to understanding moments of historical rupture but also to how political allegiances are formed in contemporary democracies. The literature on polarization has highlighted partisan sorting, media biases, divergent economic situations, and cultural divisions as key drivers of political alignment (e.g., \citealp{Mason2018UncivilAgreement}; \citealp{prior2013media}; \citealp{autor2020importing}; \citealp{hochschild2016strangers}). Less is known about the role of peers and social networks in shaping political loyalties in polarized settings. While research has shown that peers matter for outcomes such as education, consumption, health, and careers,\footnote{We cannot include a comprehensive review of the literature here, but such overviews are available in  \citet{benhabib2010handbook}, \citet{brechwald2011beyond}, \citet{aral2017exercise}, \citet{mcgloin2019peer}, \citet{giletta2021meta}, and \citet{boucher2024toward}.} there is little causal evidence as to whether peers influence high-stakes political and/or life decisions, especially when these decisions come with severe personal risks and moral conflicts.

The American Civil War provides an opportunity to examine this question. West Point cadets decisions as to which side to fight for 
pitted personal backgrounds, political-economic interests, and regional loyalties against professional ties and peer influences. Although West Pointers played pivotal roles in the war, systematic empirical analysis of the factors influencing their allegiances remains limited.\footnote{There exists a rich historical literature on West Pointers in the Civil War (e.g., \citealp{patterson2002rebels}). A recent political science study, \citet{white2024rebel}, highlights the importance of economic interests. See also additional references below.}
This setting offers two key advantages. First, the quasi-random composition of cadet cohorts generates exogenous variation in exposure to peers from different regions. This allows us to separate peer effects from confounding factors (e.g., individuals typically self-select into like-minded groups) that typically make it difficult to disentangle network influence from shared beliefs or communal forces (see \citealp{angrist2014perils}). Quasi-randomness arose from institutional features: although admissions broadly tracked the geographic distribution of congressional districts, annual recruitment varied substantially due to decentralized nomination procedures and qualification exams (e.g., \citealp{park1840westpoint}). 
Second, West Point was the nation's premier military academy, where cadets from Northern (Free) and Southern (Slave) lived and trained together, and forged close bonds before the war. This environment allowed for peer influence. Historians have long noted that ``group cohesion'' and camaraderie motivated enlistment and fighting, but the evidence has been largely anecdotal, drawn from letters and diaries (e.g., \citealp{mcpherson1997cause, mcpherson2003battle, siebold2007essence}). Whether peers swayed cadets' choices of which army to join, or whether local political and economic pressures outweighed such influence, has been an open question.

To answer this, we digitize detailed biographical data on more than 1,600 West Point cadets from 1820 to 1860 and exploit quasi-random variation in peer composition. We classify states as Free or Slave based on whether slavery was legal in 1860 (corresponding to a slave population share exceeding 1\%), and confirm robustness with alternative thresholds. The share of peers from Free States fluctuated annually without clear trends, creating plausibly exogenous exposure to Northern peers (see Section~\ref{sec_descriptive} for institutional details). This variation is not systematically correlated with individual backgrounds or home-state characteristics, whether in the full sample or when restricting to cadets who later fought in the war, supporting its interpretation as quasi-random. In the main text, we focus on the wartime choices of cadets who served in the conflict. The Supplemental Appendix presents additional robustness checks using the full cadet sample.

Our primary finding is that there is significant and large peer effect on cadet's decisions.  A higher proportion of classmates from Free States significantly increased the likelihood that cadets from Slave States joined the Union Army. The effect is sizable: a one-standard deviation increase in the proportion of peers from Free States (an increase of roughly 5 out of 40 peers) raised the probability of joining the Union by 5.4 percentage points, or 15\% of the mean. In contrast, nearly all cadets from Free States joined the Union regardless of cohort composition. The difference suggests a conflict between nationalism and sectionalism faced by cadets from Slave States, who were more susceptible to peer influence. We formalize this intuition with a simple framework that illustrates why there are asymmetric peer effects between cadets from Slave and Free States.

We also find that peer effects vary with political-economic background, which shapes the tradeoff inherent in the allegiance decision. 
Using slave-holding data from the 1860 Census and genealogical records from Ancestry.com and Familysearch.com, we find that among cadets from Slave States, stronger ties to slavery, proxied by a very high slave population shares in one's home state (or county) or by personal slave ownership, essentially eradicate peer influence.  The peer effect is very large in regions and for cadets less tied to slavery.  
 Moreover, slave population share emerges as a more significant proxy for political-economic background with regards to cadet's decisions than alternative measures, including county-level presence of pro- or anti-slavery religious groups and county-level voting shares for pro- and anti-slavery candidates in the 1860 presidential election.

Our third set of analyses highlights the role of interaction in shaping peer influence. We find that peer effects are stronger within cohorts than across cohorts, indicating the importance of direct and contemporaneous interaction. Peer influence is also more pronounced among cadets who maintained continuous military careers between their graduation and the Civil War, compared with those who had returned to civilian life, suggesting that sustained embeddedness amplified interpersonal influence. 
We also track variation in cadets' likelihood of having served together in the Mexican–American War (1846–1848). Peer effects are stronger when cadets from Slave States had more Free-State peers with whom they shared combat experience, and weaker when their co-combatants were also from Slave States. Together, these findings provide suggestive evidence that peer influence operates through repeated/sustained interactions.

In our additional analysis, we examine heterogeneity and test the robustness of our results, including tests that account for shifting political environments, peer dropouts, those not participating in the war, appointment states, and other factors. We also trace some of implications that cadets' decisions had on their careers and lives. Because many West Point cadets in our study later emerged as significant military figures, we can examine how their wartime choices influenced their post-war trajectories. Employing both OLS and an instrumental variable strategy (where home state and peer exposure predict allegiance), we find that for those from Slave States, those who sided with the Union experienced lower military ranks in 1865 and a diminished chance of becoming generals, but also a decreased likelihood of wartime mortality. 

Our study contributes knowledge about peer effects (see \citet{benhabib2010handbook,brechwald2011beyond,mcgloin2019peer,giletta2021meta} for some overviews), by demonstrating how peer influence can shape major life decisions, particularly during moments of polarization. More broadly, it adds to the long-standing literature emphasizing the importance of ``personal influence'' in economic, political, and social behaviors (e.g., \citealp{KatzLazarsfeld1955}).
Other research specifically involving soldiers concerns the spread of political ideas (\citealp{jha2023revolutionary,grosjeanpolitical2025}), which our study complements in studying allegiance (and how it depends on economic circumstances).

The experiences of West Point cadets during the American Civil War mirror other historical contexts where military academies produced leaders for both sides of major conflicts in divided nations. For instance, the Whampoa Military Academy in the 1920s produced commanders who fought on opposing sides during the Chinese Civil War (e.g., \citealp{bai2023ideas}), and the Heroic Military Academy in Mexico trained leaders who both supported and opposed the regime during the Mexican Revolution. Such decisions are not unique to military cadets, as similar choices are faced by ordinary citizens in many divided societies. However, analyzing peer influence in these broader contexts is often more challenging. The distinctive setting of the American Civil War, where allegiances were closely tied to geography, provides an opportunity to examine peer influence with a level of identification that is difficult to achieve in other cases.

Our findings also add to the body of historical and social sciences literature that examines loyalty decisions during American Civil War. \citet{mcpherson1997cause} draws on soldiers' letters and diaries to investigate soldiers' motivations. Analyzing Union Army company data, \cite{costa2003cowards,costa2008} emphasize the importance of loyalty to fellow soldiers, which sometimes surpassed commitment to the cause when considering decisions to fight, shirk duty, or desert. Our findings suggest that both cause (the allegiance to one's home state) and comrades (peer influence) matter and that they also interact with each other. Closely related, \cite{hall2019wealth} document that slave ownership drove enlistment in the Confederacy, whereas \cite{white2024rebel} explores the allegiance of West Pointers, emphasizing economic interest by showing that one's past employment related to cash crops (i.e., sugar, indigo, rice, tobacco, and cotton) is linked to the likelihood of West Pointers joining the Confederacy.\footnote{Using the same variable on cash-crop employment, we find a similar pattern. But with only a small share of candidates having such a history, we do not find a significant interaction effect between this variable and peer influence.} Our observation regarding the influence of the slave economy corroborates these findings, though our primary focus remains on peer influence—a critical factor in loyalty decisions during pivotal historical moments.

Beyond this context, a literature has examined the determinants of conflict participation, highlighting the importance of interests (as reviewed in
\citet{blattman2010civil}) and social forces (reviewed in \citet{cantoni2024protests}).\footnote{There is also a long-standing literature in sociology recognizing the importance of personal networks in social and political movements (e.g., \citealp{mcadam1986recruitment}, \citealp{diani2003social}).} We contribute to the previous literature in terms of the variation in peer composition that identifies peer influence,  the differing ways that it works depending on background of a cadet's state, the decisions made by leaders, and the choice between competing ideologies.

Finally, our study complements a literature on  leadership (e.g., \citealp{jones2005leaders}; \citealp{acemoglu2015history}; \citealp{dippel2021leadership}; \citealp{cage2023heroes}; \citealp{bai2023web}; \citealp{ferrara2025frontline}). Understanding how leaders develop their ideas and loyalties can have  political and economic implications. Our findings highlight the significance of peer influence in shaping future leaders' decisions during crises, demonstrating how peer influence drives individual choices that can have historical implications.

\section{The Data} \label{sec_data}

We now describe how the data are constructed. Additional details appear in the Materials and Methods Section below and the Supplemental Appendix \ref{Dataconstruct}.

\subsection{The Cadets} \label{sec_cadets}

\noindent We focus on cadets who graduated from West Point Military Academy between 1820 and 1860.\footnote{We begin in 1820 because we observe that some cadets back to this date participated in the Civil War. We also report additional results that restrict the sample to later cohorts only.} We manually collected information from the comprehensive \textit{Biographical Register of the Officers and Graduates of the United States Military Academy} \citep{cullum1891biographical}. The records provide details such as the year of graduation, birth year, home state and appointment state,\footnote{Home state and appointment state usually overlap. We use home state in our main analysis and present additional robustness checks using appointment state to define peers in the Supplemental Appendix.} and academic rank. To ensure comparability of academic performance, we classify cadets into percentiles based on their graduation ranking. The top 10\% were assigned a score of 100, the bottom 10\% a score of 10, and so on. Additionally, the records include information on wars and battles one participated in and their military ranks, which we use to determine if one joined the Union or Confederate forces during the Civil War. We present examples of these records and our coding procedure in Supplemental Appendix \ref{A_data}.

Since the Biographical Register lacks some information about eventual military ranks and death for some cadets who joined the Confederacy, we supplement our data with sources including \textit{Rebels from West Point} \citep{patterson2002rebels}, \textit{Southern Historical Society Papers} \citep{herndon1876southern}, and \textit{Confederate Military History} \citep{evans1899confederate}. We also account for cadets who did not graduate from West Point during 1820-1860, using the \textit{Register of Graduates and Former Cadets} \citep{thayer1964register}. Additionally, we cross-reference our data with Wikipedia and epitaph data \citep{findagrave} whenever information about a cadet is accessible.

We collected additional data on cadet backgrounds including (i) a cadet's home county, which can be linked to county-level characteristics; (ii) cadet slave ownership (whether the cadet personally owned slaves in 1860); (iii) whether a cadet's father and mother originated from Free States;
and (iv) whether a cadet's wife originated from a Free State (with unmarried cadets coded as not having such a spouse). These data are obtained from Ancestry.com and FamilySearch.com, including linked data from the 1860 Census. We describe our data-linking procedure in the Supplemental Appendix \ref{A_data}.

To proxy cadets' political–economic background, we use the slave population share in a cadet's home state and home county, as well as whether the cadet personally owned slaves in 1860. The first two measures are drawn from the 1860 Census \citep{census1860}, and cadet slave ownership is obtained from Ancestry.com. We also examine a broader set of economic proxies (including log farmland value per capita, log manufactured output per capita, and the share of employment in manufacturing) as well as religious and voting measures. Across specifications, a cadet's region's slave population  and the cadet's slave ownership emerge as the most robust predictors of wartime allegiance.

Following \cite{white2024rebel}, we code cadets' employment histories and use involvement in cash-crop occupations as a proxy for material interests, allowing us to compare this measure with the slave population share. In addition, we leverage data on cadets' participation in the Mexican-American War (1846–1848) to shed light on potential channels through which peer influence may have operated.

\subsection{Geography and the Classification of States by High/Low Slave} \label{sec_geography}

We are interested in how cadets' decisions were shaped by the choices of their peers. To address the challenge of endogeneity in peer choices, we use cadets' home state as a source of exogenous variation. Specifically, we explore the impact of the proportion of cadets from Free States on the allegiance decisions of cadets from both Free and Slave States.\footnote{Our approach is consistent with the recommendation in \citet{angrist2014perils}, where ``research designs that manipulate peer characteristics in a manner unrelated to individual characteristics provide the most compelling evidence on the nature of social spillovers.''}

There were 11 states that seceded and formed the Confederate States of America by the end of 1861 (Alabama, Arkansas, Florida, Georgia, Louisiana, Mississippi, North Carolina, South Carolina, Tennessee, Texas, Virginia).  There were four states (Delaware, Kentucky, Maryland, and Missouri) that were considered Border States\footnote{The District of Columbia is generally also added to the border category, although not a state.} and that had significant internal splits in their allegiance, and in some cases even parallel governments.

In the main text we categorize states existing at the beginning of the war\footnote{West Virginia was split off from Virginia in 1863, becoming part of the Union, but we treat it as part of Virginia as of 1860.} into Slave and Free States based on whether slavery was legal in 1860, which corresponds to using a 1\% slave population share in 1860 as a threshold. In the Supplemental Appendix, we show that our findings on peer effects remain (and can become even stronger) when using higher thresholds of 5\% or 10\% of slave populations. Supplemental Appendix Table \ref{Atab_states} lists the classification of states using different thresholds.

The territories that were not yet states were under Union government control and none could secede, and none had significant slave populations.\footnote{A few had split allegiances, but overall the territories provided relatively small enlistments, and mostly for the Union.} There is only one cadet in our sample from these territories, which we exclude from our analysis. We also exclude the few cadets whose hometowns are outside the U.S.

Between 1820 and 1860, 1,656 cadets graduated from the United States Military Academy (see the complete records in Biographical Register of the Officers and Graduates of the United States Military Academy \citep{cullum1891biographical}).  17 graduates were from outside the continental United States, and one was born in Indian Territory, our analytical sample consists of 1,638 cadets. In our data, 968 and 670 cadets came from Free States and Slave States, respectively. Among those from Free States, 540 joined the armies and  91.9\% of them joined the Union. Among those from Slave States, 388 joined the armies and 36.1\% of them joined the Union. 

Supplemental Appendix Table \ref{Atab_summary} presents the summary statistics of the key variables using the baseline threshold.  Further, Supplemental Appendix Table \ref{Atab_cohort_summary} reports the cohort-level distribution of graduates by state type, along with their standard deviations, showing that there was substantial variation from cohort to cohort in the fraction of cadets from Free vs Slave States.

\section{Results}\label{sec_results}

\subsection{Descriptive Evidence} \label{sec_descriptive}

\begin{figure}[H] 
    \begin{center}
    \begin{minipage}{0.48\linewidth}
        \centering
        \vspace{3pt}
        \includegraphics[width=\textwidth,trim=1 1 1 1,clip]{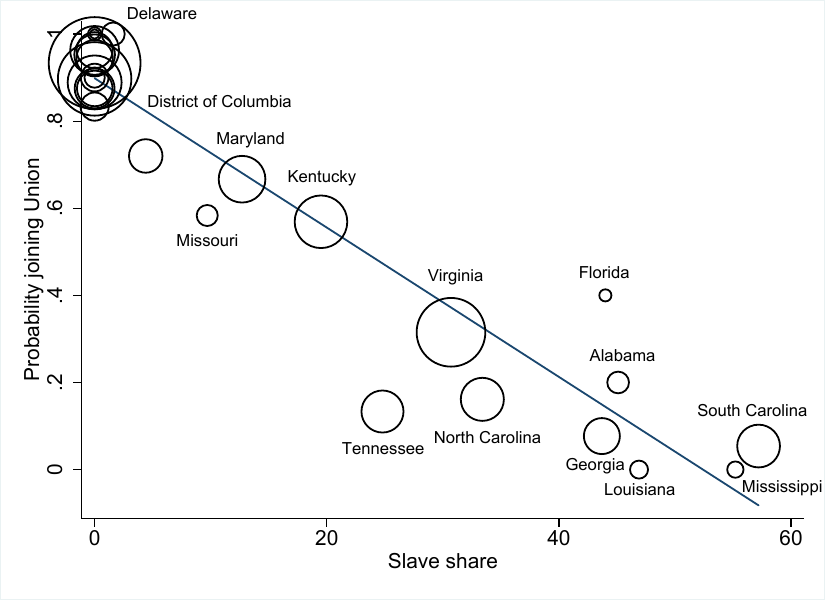}
        \scriptsize A: Probability of joining the Union: war participants
    \end{minipage}
    \hfill
    \begin{minipage}{0.48\linewidth}
        \centering
        \vspace{3pt}
        \includegraphics[width=\textwidth,trim=1 1 1 1,clip]{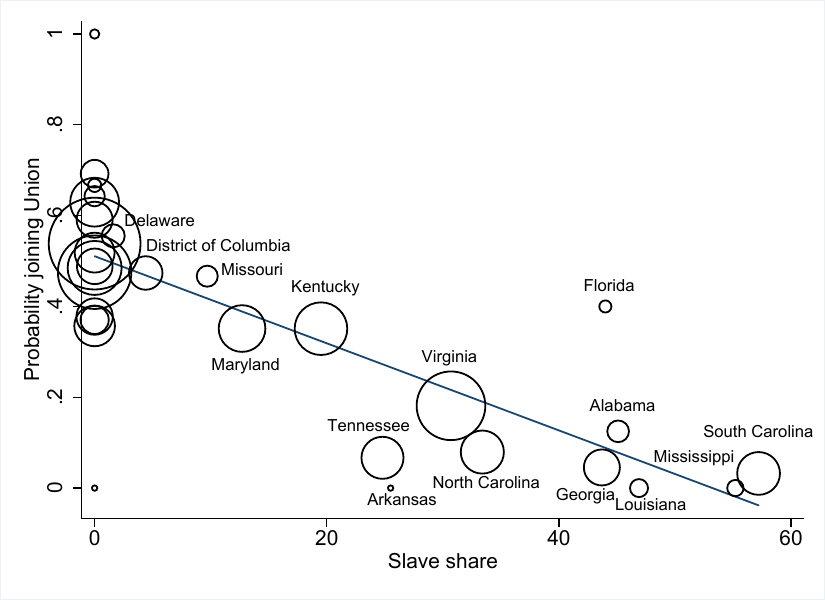}
        \scriptsize B: Probability of joining the Union: all cadets
    \end{minipage}
    \caption{Home-state slave share and the probability of joining the Union} \label{fig_home_slaveshare}
\end{center}
  \vspace{0.2cm}
\begin{spacing}{0.75}
{\footnotesize {\scriptsize \textit{Note.} This figure plots the relationship between the share of the slave population in a state and the likelihood that cadets from that state joined the Union army. Panel A concentrates on the states that participated in the war, while Panel B uses all cadets, including those who did not participate in the war. The circle's size represents the number of West Point cadets hailing from each state.}}
\end{spacing}

\end{figure}

\paragraph{Importance of Home State}
First, we establish that a cadet's home state serves as an effective representation of regional norms and preferences, as evidenced by the considerable impact of the state's slave population percentage on the cadets' loyalty. Figure \ref{fig_home_slaveshare} illustrates that the proportion of a state's slave population is strongly and negatively related to the likelihood of joining the Union, regardless of whether we focus on war participants or all cadets. 

In general, the slave population serves as a strong predictor for joining the Union compared to the Confederacy, but it has no (significant) correlation with the decision to participate in the war (Supplemental Appendix Figure \ref{fig_byslave_Share_all}). Thus, we focus on war participants in our analysis.

\paragraph{Relevance of Peers}

\begin{figure}[ptb]
\caption{Relevance of Peers: Motivating Evidence}\label{fig_correlationscatter}
\begin{center}
  \subfloat[Free-State peers and allegiances for two groups of cadets] {\includegraphics[width=0.8\textwidth,trim=10 10 10 10,clip]{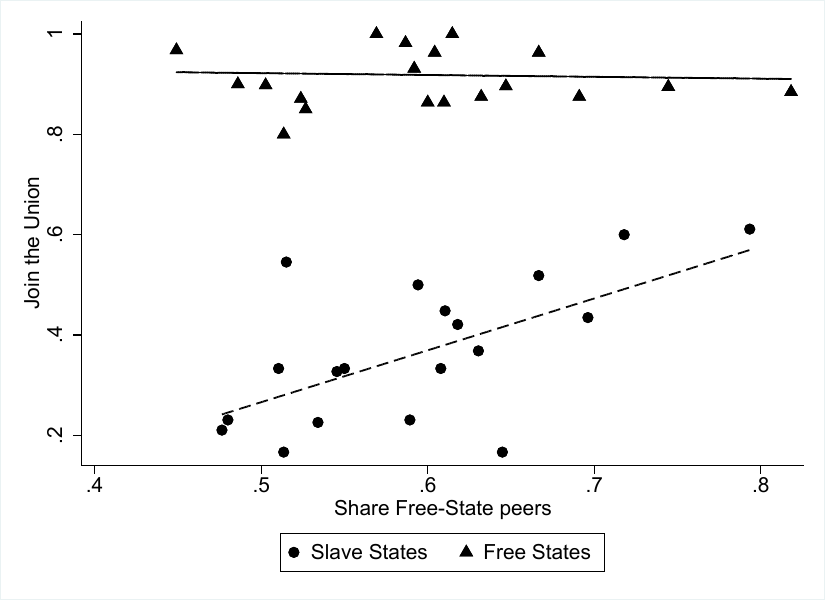}} \\
		\subfloat[Peer composition over time]{\includegraphics[width=0.8\textwidth,trim=10 10 10 10,clip]{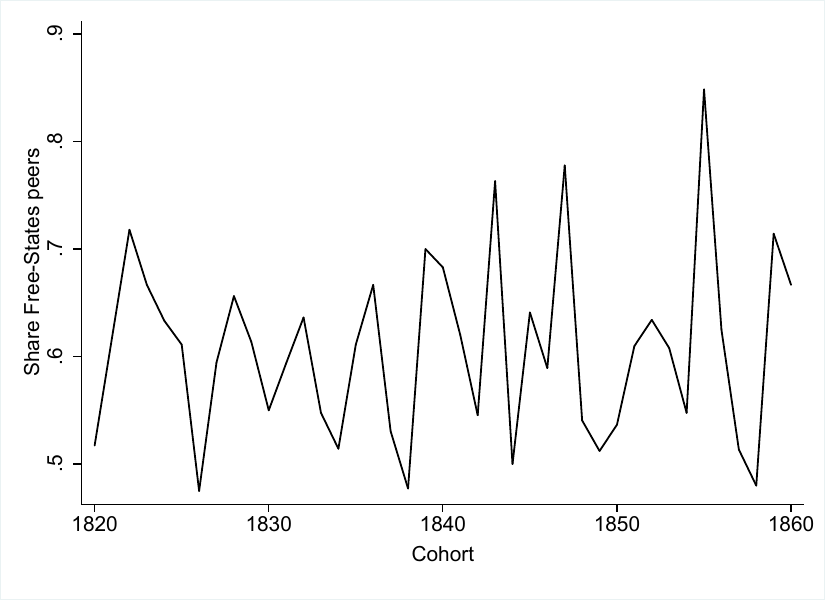}}
\end{center}
  \vspace{0.2cm}
  \begin{spacing}{0.75}
   {\footnotesize {\scriptsize \textit{Note.} Panel (a) shows the  relationship between the share of Free-State peers and the probability of joining the Union, depicted separately for war-participating cadets from Free and Slave States. Peer influence appears stronger among the latter. Panel (b) presents the time-series of peer composition, showing substantial variation over short periods of time.
   }}
  \end{spacing}
\end{figure}

In Figure \ref{fig_correlationscatter}(a), we illustrate the correlations between the proportion of free-state peers in each cohort and the likelihood of Union affiliation for cadets from both slave and Free States. The figure reveals a distinctly positive relationship for cadets from Slave States, suggesting peer influence when facing a conflict between nationalism and sectionalism. Conversely, nearly all cadets from Free States joined the Union, which aligns with their lower tendency to experience a conflict.\footnote{Of the 44 that joined the confederacy, 11 were from New York and 10 from Pennsylvania.  The 44 were from slightly earlier cohorts than average and had slightly higher class rank than average.}

\paragraph{What Drives Variation in Peer Composition?}

As Figure \ref{fig_correlationscatter}(b) shows, there was substantial year-to-year variation in the peer composition of West Point cadets, which enables our main identification of peer effects. This variation stems from two key features of the Academy's admissions process (e.g., \citealp{park1840westpoint}).

First, admissions were designed to achieve geographic proportionality: the number of cadets from each state was intended to reflect its congressional representation, and, roughly, its population.
The admission of cadets required a nomination from a member of the House of Representatives (which now has expanded to also include senators).  The slot constraints (some of which went unfilled while others were over-subscribed) provides for substantial randomness in home-state composition from cohort to cohort.
We find that logged state population in 1820 alone can explain 24\% of the variation in the number of admitted cadets across states.

Second, despite this long-run proportionality, annual admissions fluctuated considerably. Several factors contributed to this. Not all congressional districts submitted nominations each year—many representatives failed to exercise their appointment privilege or were unable to identify qualified candidates. Even when nominations were made, candidates had to pass West Point's entrance exams; and if a nominee failed, the slot could go unfilled or be reassigned through an at-large appointment. Furthermore, prior to 1843, presidential discretion added an additional layer of unpredictability to the process.\footnote{In Supplemental Appendix Table \ref{tab_validation}, we examine the correlations between $Peer_{i,t}$ and $X_{i,s}$. Once the time trend is accounted for, there is no strong relationship between personal characteristics and peer composition. Moreover, the balance tests hold whether we restrict the sample to cadets who joined the war or include all cadets.  Thus, we do not see any observable systematic pattern in the peer composition from cohort to cohort, and cannot reject the identification assumption.}

These features were interrelated. Because of the underlying commitment to proportionality, a higher number of cadets from a given state in one year often implied fewer admissions from that state in subsequent years.

\subsection{Statistical Analysis}

Motivated by the descriptive evidence, we estimate the impact of peer influence via the following OLS baseline specification:
\begin{equation}
Union_{i,s,t} = \beta_0 + \beta_1 Peer_{i,t} + \gamma X_{i,s} + \alpha_s + \theta t + \epsilon_{i,s,t},
\end{equation}
where $Union_{i,s,t}$ denotes whether cadet $i$ from state $s$ and cohort $t$ joined the Union. $Peer_{i,t}$ represents the proportion of peers from Free States.\footnote{Note that our specification avoids the exclusion bias noted by \cite{caeyers2024exclusion}, since we are separating regressions for Slave and Free States and working with exogenous characteristics.} $\alpha_s$ represents state fixed effects.

$X_{i,s}$ includes a range of personal characteristics. Our main analysis focuses on variables available for all cadets recorded in \citep{cullum1891biographical}: age in 1860, academic rank at West Point, and the slave population share in the cadet's home state. When home-state fixed effects are included, state-level characteristics are naturally absorbed. In addition, we consider alternative measures of political-economic background, including whether a cadet owned slaves in 1860, and indicators for whether the cadet's father, mother, or wife originated from Free States. These variables are constructed for the subset of cadets for whose information is available from Ancestry.com and FamilySearch.com.

We incorporate a linear time trend $t$ to account for potential temporal trends in Union enlistment. In addition, we report estimates that incorporate five-year bin fixed effects, which help absorb shifts in the political environment. To address potential serial correlation, we compute bootstrapped standard errors based on 400 resampling iterations. 

Our specification examines peer effects based on exogenous peer characteristics and not endogenous peer choices (\cite{manski1993identification}).  Given that cadets' decisions were made some years after graduation, it is less likely that there was direct coordination on decisions rather than lasting impacts of interactions from their formative years.  Nonetheless, it is possible that there is correlation in their decisions, and thus for robustness, we also re-estimate the model using standard errors clustered at the cohort level as well as bootstrapped standard errors clustered at the cohort level (Supplemental Appendix Table \ref{tab_different_SE}).
We also directly examine residual dependence across peer groups by estimating intra-class correlation coefficients (ICC in Supplemental Appendix Table \ref{tab_different_SE}) for the baseline residuals within cohorts. The results show negligible within-cohort residual correlation, providing no evidence of jointly determined wartime allegiance decisions.

As mentioned, we focus on those who joined the armies in our baseline analysis, which is the main margin of peer influence. In addition, we also employ a multinomial logit specification where we consider three outcomes: joining the Union, joining the Confederacy, and not joining the war.


\subsection{The Impact of Peers on Cadets' Allegiance}

\paragraph{Baseline Estimates}
Table \ref{tab_baseline_10pct} displays our main baseline results, indicating that peer influence matters for individuals from Slave States. Column (1) illustrates the raw correlation between the proportion of Free-State peers and the likelihood of joining the Union. Column (2) incorporates personal and state characteristics, while Column (3) further includes state fixed effects. To facilitate the interpretation of coefficients, we present the coefficient for a one standard deviation change in peer exposure. In Column (4), we consider the subset of cadets with additional individual background information. Across all specifications, we observe a strong  peer influence for individuals from Slave States.

The peer effect is sizable:  a one standard deviation increase in of free state peers is roughly 5.6 more out of 40 cadets\footnote{See Appendix Supplement Table \ref{Atab_cohort_summary}.} in the cohort being from Free States and leads to an 15\% increase in the frequency of joining the union (mean 0.36). This is a conservative estimate. When using alternative thresholds for Slave States (5\% and 10\%), we find that peer effect raises the probability of Slave-State cadets joining the union by 21\% (see Supplemental Appendix Table \ref{Atab_thresholds2}).
The effect is above 30\% when we examine cadets who were not slave owners in 1860 (discussed later).

The standard errors are bootstrapped with 400 resampling iterations. In Supplemental Appendix Table \ref{tab_different_SE}, we also report standard errors clustered at the cohort level as well as those bootstrapped at the cohort level. The estimated coefficients on peer influence remain unchanged, while the corresponding p-values are generally smaller. To remain conservative, we rely on the larger standard errors in our main analysis.

\begin{table}[ptb]
		
   \caption{ \centering Peer composition and allegiance choice}\label{tab_baseline_10pct}
		
    \begin{center}
			
			\footnotesize 
    \def\sym#1{\ifmmode^{#1}\else\(^{#1}\)\fi}
    \resizebox{\textwidth}{!}{
        \begin{tabular}{l*{8}{c}}
            \hline\hline
            Dependent var
            &\multicolumn{8}{c}{Join the Union: War Participants}
            \\\cmidrule(lr){2-9}
            &\multicolumn{4}{c}{Slave-State Cadets} 
            &\multicolumn{4}{c}{Free-State Cadets}
            \\\cmidrule(lr){2-5}\cmidrule(lr){6-9}
            &\multicolumn{1}{c}{(1)} 
            &\multicolumn{1}{c}{(2)} 
            &\multicolumn{1}{c}{(3)} 
            &\multicolumn{1}{c}{(4)}
            &\multicolumn{1}{c}{(5)} 
            &\multicolumn{1}{c}{(6)}
            &\multicolumn{1}{c}{(7)} 
            &\multicolumn{1}{c}{(8)}\\
            \hline
            Share Free-State Peers (sd)    &    0.080\sym{***}&    0.058\sym{***}&    0.054\sym{***}&    0.056\sym{**} &   -0.003         &   -0.005         &   -0.006         &   -0.005         \\
                &  (0.023)         &  (0.020)         &  (0.020)         &  (0.022)         &  (0.013)         &  (0.013)         &  (0.014)         &  (0.014)         \\
            Age in 1860        &                  &   -0.012         &   -0.005         &    0.004         &                  &    0.006         &    0.004         &   -0.001         \\
                &                  &  (0.015)         &  (0.016)         &  (0.016)         &                  &  (0.007)         &  (0.007)         &  (0.008)         \\
            Class Rank      &                  &    0.001         &    0.001         &    0.002\sym{**} &                  &   -0.001\sym{**} &   -0.001\sym{**} &   -0.001\sym{***}\\
                &                  &  (0.001)         &  (0.001)         &  (0.001)         &                  &  (0.000)         &  (0.000)         &  (0.000)         \\
            Slave Pop. Share (sd)&                  &   -0.216\sym{***}&                  &                  &                  &   -0.007         &                  &                  \\
                &                  &  (0.019)         &                  &                  &                  &  (0.016)         &                  &                  \\
            Cohort    &                  &   -0.013         &   -0.005         &    0.002         &                  &    0.008         &    0.007         &    0.003         \\
                &                  &  (0.015)         &  (0.015)         &  (0.016)         &                  &  (0.007)         &  (0.007)         &  (0.008)         \\
            Cadet Slave Ownership   &                  &                  &                  &   -0.276\sym{***}&                  &                  &                  &            \\
                &                  &                  &                  &  (0.046)         &                  &                  &                  &          \\
Free-State Father &                  &                  &                  &    0.090         &                  &                  &                  &    0.019         \\
                &                  &                  &                  &  (0.070)         &                  &                  &                  &  (0.061)         \\
Free-State Mother &                  &                  &                  &    0.077         &                  &                  &                  &    0.034         \\
                &                  &                  &                  &  (0.075)         &                  &                  &                  &  (0.060)         \\
Free-State Wife &                  &                  &                  &    0.061         &                  &                  &                  &    0.098\sym{***}\\
                &                  &                  &                  &  (0.057)         &                  &                  &                  &  (0.028)         \\

            State FEs &N&N&Y&Y&N&N&Y&Y\\
        \hline
        Dependent var. mean &0.361&0.361&0.361&0.355&0.919&0.919&0.919&0.930\\        
Observations    &      388         &      388         &      388         &      324         &      540         &      540         &      540         &      456         \\
R-squared       &    0.028         &    0.234         &    0.287         &    0.363         &    0.000         &    0.017         &    0.030         &    0.067         \\
        \hline\hline
        \end{tabular}
}

	\end{center}
\par
\vspace{0.2cm}
\begin{spacing}{0.75}
{\footnotesize {\scriptsize \textit{Note.} This table presents the impact of the fraction of peers from Free States in a cadet's cohort on that cadet's decision to join the Union army. Columns (1)-(4) focus on cadets from Slave States (i.e., with a slave population higher than 1\%) and Columns (5)-(8) on those from Free States (i.e., with a slave population lower than 1\%). The variable \textit{Free-State father} and \textit{Free-State mother} equal 1 if a parent is from a Free State, and 0 otherwise. A subset of observations have missing information on parental origin.  \textit{Free-State wife} equals 1 if the wife is from a Free State and 0 otherwise. If the wife's information is missing, we assign a value of 0. The standard errors presented in the parentheses are obtained through bootstrapping with 400 resampling iterations. ***$p<0.01$, **$p<0.05$, *$p<0.1$.}}
\end{spacing}

\end{table}

We include cadets who graduated between 1820 and 1860 in our baseline analysis because cadets dating back to the early cohorts (from 1820) participated in the Civil War, often serving in officer or administrative roles. As a robustness check, we re-estimate our baseline specifications after excluding those aged 60 (or 50) and above; the estimated peer effects remain quantitatively similar and statistically robust. The results are reported in Appendix Table \ref{tab_different_age}.

Among the personal background variables, the share of the slave population in a cadet's home state is strongly and negatively associated with joining the Union, underscoring the role of political-economic factors in allegiance decisions. Supplemental Appendix Table \ref{Atab_background_variables} presents correlations between the probability of joining the Union and additional state-level economic proxies, including logged farmland value per capita, logged manufactured product value per capita, and the share of manufacturing employment. Notably, incorporating these proxies does not alter our primary findings on peer effects. Furthermore, the share of the slave population emerges as the most robust economic predictor of war allegiance. All these economic variables are effectively accounted for when controlling for state fixed effects.

The individual-level cadet slave ownership is also a strong predictor of allegiance. As expected, it is negatively correlated with the probability of joining the Union.

In contrast to the strong peer influence observed among cadets from Slave States, we find no comparable peer effect among cadets from Free States, consistent with Figure \ref{fig_correlationscatter}. These results are reported in Columns (5)–(8). 

Next, we interpret this asymmetric peer effect in a simple framework.

\paragraph{Interpreting the Asymmetric Peer Effects}

The empirical asymmetry in peer influence---namely that cadets from Slave States respond to peer composition while cadets from Free States do not---can be understood through a simple choice framework.

Normalize the value of joining the Confederacy to zero. A cadet joins the Union if the net utility from doing so is positive. This utility consists of three components. First, there is a default institutional allegiance associated with West Point and service in the U.S. Army. Let this term be $V>0$. Second, cadets from Slave or Border States may have material or ideological interests in the Confederacy—denoted $S_i\ge 0$, with $S_i=0$ for cadets from Free States. Third, peer interactions depend on the fraction of a cadet's cohort originating from Free States, $F_i$. We capture peer influence via $\beta(F_i - 1/2)$, which increases (decreases) the utility of joining the Union when Free-State peers are in the majority (minority).

Total utility of joining the Union for cadet $i$ is
\[
U_i = V - S_i + \beta(F_i - 1/2) + \varepsilon_i,
\]
where \( \varepsilon_i \) is an idiosyncratic shock with distribution \( G(\cdot) \).

Cadets thus join the Union as long as
\[
\varepsilon_i > S_i -V - \beta(F_i - 1/2).
\]
Thus, the probability of joining the Union is for cadet $i$ is:
\begin{equation}\label{Choice}
P_i= 1-G (S_i - V -\beta(F_i - 1/2)).
\end{equation}

For Free-State cadets $S_i$ is 0 and so this probability is close to 1.
For cadets heavily exposed to slavery (either from a region heavily dependent on slavery, or personally owning slaves), $S_i$ is large and this probability is close to 0.  
It is for intermediate values of $S_i-V$
that this varies substantially as a function of $\beta F_i$.

For example, if $G$ is a uniform distribution on $[0,1]$, then 
(\ref{Choice}) becomes 
\[
P_i={\rm min}\left[1, \left( 1-\beta/2+ V-S_i  + \beta F_i \right)^+ \right],
\]
which is a constant 1 for low $S_i$ and $V>1-\beta/2$ (Free-State cadets), consistent with what we have found above.
It is a constant 0 for high $S_i$, and has a slope of $\beta$ in $F_i$ for intermediate values of $S_i$.   
Below, we verify this interaction of the peer effect with the magnitude of the slave interest of cadets ($S_i$).

More generally, differentiating \eqref{Choice} yields
\[
\frac{\partial P_i}{\partial F_i}
= \beta\, g\!\left(S_i - V - \beta(F_i - 1/2)\right),
\]
where \( g(\cdot) \) is the density of \( G(\cdot) \).

For Free-State cadets (\(S_i=0\)) and sufficiently large \(V\),  this is close to zero because choices are far from the margin where the density is small, consistent with what we found above. 
Moreover, among Slave-State cadets, a high $S_i$ mitigates peer effects. When $S_i$ is very large, $\frac{\partial P_i}{\partial F_i}$ is also far from the margin and close to zero. 
With this in hand, we examine how peer influence varies with various proxies for \(S_i\).

\paragraph{Variations in Peer Effects with Political-Economic Backgrounds}

We use several proxies to measure cadets' material and ideological interests in supporting the Confederacy.
Our main focus is on exposure to and personal investment in slavery: the slave population share in a cadet's home state, the slave population share in his home county, and whether he owned slaves in 1860. In Panel A of Table \ref{Atab_3groups}, we interact our peer measure with the state-level slave population share and with cadet slave ownership. The peer effect is smaller where slavery was more prevalent, and likewise among cadets who owned slaves.
Deeper entanglement in the slave economy reduced responsiveness to peers.  Correspondingly, once we control for this the magnitude of the peer effect among the remaining slave-state cadets becomes even larger.  For example, among slave-state cadets who did not personally own slaves, we see that the coefficient is now .076, more than a third increase over the coefficient not controlling for this.

We further divide the Slave States into groups by the relative slave populations.  It turns out that the peer influence operates entirely on cadets from the range of Slave States that have slave populations below one third of the population. In Panel B of Table \ref{Atab_3groups}, we break states into three groups:  those without slaves, those with slaves up to one third of the population, and those with slave populations above one third.   We now see that the peer effect comes entirely from this middle group of states, and in fact the estimated coefficient and significance are larger than that from the previous table. These findings suggest that in High-Slavery States, economic imperatives and entrenched social norms overrode peer influence, muting its effect. This is consistent with a very high $S_i$ in equation (\ref{Choice}).

\begin{table}[h]
		
   \caption{ \centering Peer composition and allegiance choice: Different groups}\label{Atab_3groups}   
    \begin{center}			
		\footnotesize	

    \def\sym#1{\ifmmode^{#1}\else\(^{#1}\)\fi}
\resizebox{\textwidth}{!}{
\begin{tabular}{l*{3}{c}}
\hline\hline
Dependent var &\multicolumn{3}{c}{Join the Union: War Participants}\\
\cmidrule(lr){2-4}
&\multicolumn{3}{c}{Slave Ownership by the Cadet's:}\\
 & State  & County & Self \\
\cmidrule(lr){2-4}
 &\multicolumn{1}{c}{(1)} &\multicolumn{1}{c}{(2)} &\multicolumn{1}{c}{(3)}\\
\hline
A. Interaction with Slave Variables \\
\hline
Share Free-State Peers (sd)   &    0.053\sym{***}&    0.053\sym{**} &    0.076\sym{***}\\
                              &  (0.019)         &  (0.021)         &  (0.023)         \\
Share Free-State Peers (sd) $\times$ Slave Pop. Share (State) &   -0.030\sym{*}  &                  &                  \\
                              &  (0.017)         &                  &                  \\
Share Free-State Peers (sd) $\times$ Slave Pop. Share (County) &                  &   -0.036\sym{*}  &                  \\
                              &                  &  (0.018)         &                  \\
Share Free-State Peers (sd) $\times$ Cadet Slave Ownership      &                  &                  &   -0.083\sym{**} \\
                              &                  &                  &  (0.037)         \\
Controls & Y& Y& Y\\
State FEs & Y& Y& Y\\
\hline
Dependent var.mean & 0.361 & 0.358 & 0.361  \\
Observations       & 388   & 352   & 388    \\
R-squared          & 0.291 & 0.341 & 0.357  \\
\hline
\hline
B. Different Groups of States &\multicolumn{3}{c}{Join the Union: War Participants}
            \\\cmidrule(lr){2-4}

             &    Heavy-Slave State         &    Border and Mid-Slave State   &   \quad\quad\quad Free State   \quad\quad\quad    \\
Slave Share
            &\multicolumn{1}{c}{(\textgreater33\%) Cadets}   
            &\multicolumn{1}{c}{(1\%-33\%) Cadets}
            &\multicolumn{1}{c}{($\leq$ 1\%) Cadets}
            \\\cmidrule(lr){2-4}
            &\multicolumn{1}{c}{(1)} 
            &\multicolumn{1}{c}{(2)} 
            &\multicolumn{1}{c}{(3)}  \\
            \hline   
            
            Share  Free-State Peers (sd)   &    0.010         &    0.072\sym{**} &   -0.013         \\
                &  (0.038)         &  (0.029)         &  (0.014)         \\
            Share Border-State Peers (sd)   &   -0.044         &    0.009         &   -0.018         \\
                &  (0.032)         &  (0.033)         &  (0.017)         \\
            Controls & Y& Y& Y\\
            State FEs & Y& Y& Y\\
            \hline
            Dependent var.mean & 0.107 & 0.477 & 0.919  \\
            Observations    &      122         &      266         &      540         \\
R-squared       &    0.149         &    0.205         &    0.033         \\
            \hline\hline
        \end{tabular}
}
	\end{center}
\par
\vspace{0.2cm}
\begin{spacing}{0.75}
{\footnotesize {\scriptsize \textit{Note.} The sample for this analysis consists of all war-participating cadets. The control variables consist of \textit{Age in 1860}, \textit{Class Rank}, and \textit{Cohort}. For Panel A, Columns (1)–(3) include controls for state-level \textit{Slave Population Share (sd)}, county-level \textit{Slave Population Share (sd)}, and \textit{Cadet Slave Ownership}, respectively. The standard errors reported in the parentheses are derived from bootstrapping with 400 resampling iterations. ***$p<0.01$, **$p<0.05$, *$p<0.1$.}}
\end{spacing}
\end{table}

Religious and political preferences could be relevant and are also correlated with slave population share \citep{fogel1994without}. To explore this potential influence, we incorporate county-level measures of religion and voting behavior, drawing on census data on church seats by denomination and historical election returns.
We focus this analysis on Slave States, as those are where the peer effects operated.

For voting, we use the vote share for the Southern Democratic candidate (Breckinridge) as a measure of local political support for slavery, and the vote share for the Republican candidate (Lincoln) in Slave States as those either opposing or being less actively for slavery.  For religion, we include the presence of various churches that tended to be either more or less supportive of slavery in 1860 (see the details with Appendix Table \ref{A_church_vote}).

As shown in Appendix Table \ref{A_church_vote}, among the additional variables, the presence of various churches does not have a significant impact on the probability of joining the Union, while the Lincoln vote share (Lincoln) is positively associated with it. 
We also see that including these variables and their interactions with peer exposure does not alter our main findings concerning peer influence among cadets from Slave States nor the interaction of the peer effect  with slave population share. 
We note that the absence of a significant interaction between peer influence and religion may reflect the imprecision of our proxy for underlying religious attitudes, limited statistical power, or the fact that many denominations were either internally divided or supported slavery in slave states \citep{wigger2001taking, snay1993gospel}.

\subsection{Shared Experience and Peer Influence}

Our baseline results indicate that peers shaped individual decisions at a pivotal moment. To shed light on the mechanisms behind this influence, we focus on three complementary dimensions of interaction. First, we test whether peer effects are stronger within a cadet's own cohort compared to across adjacent classes. Second, we distinguish between cadets who remained in continuous military service and those who temporarily exited the Army before the Civil War, allowing us to gauge whether sustained professional identity amplified peer influence. Third, we assess whether the strength of influence depends on shared military experience: in particular joint participation in the Mexican-American War. Together, they reveal the importance of shared experience behind peer influence.

\paragraph{Influence by Cohort}

To examine how cohort exposure matters, we analyze peers by grouping them into cohorts and display the coefficients for peers within the same cohort, as well as for cohorts from t-1 to t-3 and t+1 to t+3, in Supplemental Appendix Figure \ref{fig_bycohort}. The coefficients in the figure are obtained from a single regression that includes all controls from our baseline analysis. The result reveals that peer influence is most important within the same cohort. In contrast, although there is some degree of influence from earlier cohorts (older cadets), future cohorts do not show comparable levels of influence (younger cadets).

\paragraph{Non-continuous vs. Continuous Military Service}

Distinguishing between cadets who had returned to civilian life between graduation and 1860 and those who remained professional soldiers during that time
can provide further understanding of the operation of peer effects.

To do this, we manually reviewed military service records and coded whether and when individuals exited the regular army prior to the Civil War.
A key complication is that all cadets who joined the Confederacy were required to resign from the U.S. Army, so treating any exit as a return to civilian life would mechanically misclassify the Confederates. To avoid this, we conservatively classify a cadet as having returned to civilian life only if he exited the regular army and remained out of military service for more than one year before the start of the war.
Using this definition, we split the sample into cadets with non-continuous versus continuous military service. 

As shown in Appendix Table \ref{Atab_continuous_military}, peer exposure is positive in both groups but is substantially larger and statistically significant only among cadets with continuous military service, suggesting an important role for shared professional military experience.

\paragraph{Relevance of Co-fighting in the Mexican-American War}

A significant share---30.5\%---of cadets in our dataset participated in the Mexican-American War (1846–1848), offering a valuable context to examine how shared experiences shape peer influence. For each cadet, we compute the proportion of peers who also fought in the war, distinguishing between those from slave and Free States.

If shared wartime experience intensifies peer effects, we would expect our core finding on peer influence to be stronger when a cadet had more co-fighting peers from Free States, and weaker when more such peers were from Slave States. This is precisely what we observe (see Table \ref{Atab_Mexican}), focusing on cohorts between 1820 and 1845. Although the limited sample size prevents precise estimation of this interaction effect, the magnitude is economically meaningful, suggesting that shared combat experience enhances the intensity of peer influence.

\begin{table}[h]
		
   \caption{ \centering Peer composition and allegiance choice: Mexican-American War}\label{Atab_Mexican}
		
    \begin{center}
			
			\footnotesize 
    \def\sym#1{\ifmmode^{#1}\else\(^{#1}\)\fi}
    \resizebox{\textwidth}{!}{
        \begin{tabular}{l*{6}{c}}
            \hline\hline
            Dependent var
            &\multicolumn{6}{c}{Join the Union: War Participants}
            \\\cmidrule(lr){2-7}
            &\multicolumn{6}{c}{Slave States} 
            \\\cmidrule(lr){2-7}
            &\multicolumn{3}{c}{Slave-State Cadets (\textgreater1\%)} 
            &\multicolumn{3}{c}{Slave-State Cadets (1\%-33\%)}
            \\\cmidrule(lr){2-4}\cmidrule(lr){5-7}
            &\multicolumn{1}{c}{(1)} 
            &\multicolumn{1}{c}{(2)} 
            &\multicolumn{1}{c}{(3)} 
            &\multicolumn{1}{c}{(4)}
            &\multicolumn{1}{c}{(5)} 
            &\multicolumn{1}{c}{(6)}
             \\
\hline
            Sh. Free-State Peers join Mex.-Am. War $\times$ Sh. Free-State Peers  &    0.160\sym{*}  &    0.117         &    0.109         &    0.220\sym{**} &    0.201\sym{*}  &    0.212\sym{*}  \\
                &  (0.089)         &  (0.086)         &  (0.104)         &  (0.110)         &  (0.109)         &  (0.126)         \\
            Sh. Slave-State Peers join Mex.-Am. War  $\times$ Sh. Free-State Peers  &   -0.114         &   -0.084         &   -0.068         &   -0.172         &   -0.155         &   -0.156         \\
                &  (0.082)         &  (0.080)         &  (0.092)         &  (0.107)         &  (0.106)         &  (0.117)         \\
            Share Free-State Peers (sd)       &    0.059         &    0.044         &    0.041         &    0.050         &    0.034         &    0.025         \\
                &  (0.036)         &  (0.033)         &  (0.035)         &  (0.045)         &  (0.046)         &  (0.048)         \\
            Share Free-State Peers join Mex.-Am. War (sd)     &   -0.109         &   -0.066         &   -0.058         &   -0.156         &   -0.117         &   -0.148         \\
                &  (0.086)         &  (0.079)         &  (0.095)         &  (0.104)         &  (0.106)         &  (0.120)         \\
            Share Slave-State Peers join Mex.-Am. War (sd)     &    0.091         &    0.054         &    0.058         &    0.140         &    0.113         &    0.147         \\
                &  (0.084)         &  (0.079)         &  (0.092)         &  (0.099)         &  (0.099)         &  (0.109)         \\
                
            Controls &N&Y&Y&N&Y&Y\\
            State FEs &N&N&Y&N&N&Y\\
        \hline
        Dependent var. mean &0.402&0.402&0.402&0.515&0.515&0.515\\
        Observations    &      184         &      184         &      184         &      136         &      136         &      136         \\
        R-squared       &    0.035         &    0.264         &    0.323         &    0.047         &    0.163         &    0.233         \\
        \hline\hline
        \end{tabular}
}

	\end{center}
\par
\vspace{0.2cm}
\begin{spacing}{0.75}
{\footnotesize {\scriptsize \textit{Note.} Our sample for this table is restricted to cadets from Slave States whose cohorts fall between 1820 and 1845. It reports the impact of the fraction of peers from Free States in a cadet's cohort on that cadet's decision to join the Union army. \textit{Share Free-State peers join Mexican-American War}: The proportion of Free-State peers participating in Mexican-American war with a cadets. If the cadet did not participate in the Mexican-American War, the proportion is 0. \textit{Share Slave-State peers joining the Mexican-American war}: The proportion of Slave-State peers participating in Mexican-American War with themselves. If the cadet did not participate in the Mexican-American War, the proportion is 0. The control variables consist of \textit{Age in 1860}, \textit{Class Rank}, state level \textit{Slave Population Share (sd)} and \textit{Cohort}. The standard errors presented in the parentheses are obtained through bootstrapping with 400 resampling iterations. ***$p<0.01$, **$p<0.05$, *$p<0.1$.}}
\end{spacing}
\end{table}

\subsection{Additional Analyses}

We conduct six supplementary analyses to account for (1) whether the effect of peers on cadets could vary with shifting political environments, (2) potential differences between peers who graduated and those who dropped out, (3) whether non-participants in the war affect the results, (4) whether cash-crop employment shapes peer influence, and (5) whether defining peers by appointment state rather than home state matters. Finally, in our exercise (6), we examine post-war career outcomes to trace some consequences of cadets' choices. 

\paragraph{Shifting Political Environments}

Although the variation in the composition of cadets is fairly random across time, there is a possibility that it interacts with changes in the political environment or other factors that could impact cadets' choices. We conduct two analyses to account for such possibilities.

First, we divide the sample into periods before and after 1850.  Although the issue of slavery was highly contentious throughout the period we study, it was particularly intense in the 1850s. In an effort to avert civil war, Congress passed the Compromise of 1850, which included the Fugitive Slave Law that obligated law enforcement nationwide to assist in capturing alleged runaway slaves, and many in the north resisted its enforcement \citep{mcpherson2003battle}. Against this backdrop, we use 1850 as a dividing point to assess whether peer influence experienced notable changes before and after that year. As illustrated in Supplemental Appendix Table \ref{Atab_1850}, peer influence persisted over time, with similar effects on either side of this divide. 

Second, we augment our baseline specification with fixed effects at five-year intervals. This approach accounts for unobserved shocks or gradual shifts in the political environment that may coincide with changes in cadets' decisions. As reported in Supplemental Appendix Table \ref{tab_congressional_rotation}, the estimated effects remain highly consistent with our baseline results.

Taken together, these analyses indicate that, despite potentially evolving political circumstances, peer influence remained a significant factor in shaping cadets' choices.

\paragraph{Graduate vs. Dropout Peer Influence}

We focus on graduate peer influence in our main analysis as they spent four years together at West Point. In Supplemental Appendix Table \ref{Atab_dropout}, we consider dropouts and calculate peer influence based on them as a comparison. Our estimates are robust to including dropouts in our calculation of peers. However, when separating graduating peers from dropout peers, we find that it is the graduating peers rather than the dropout peers that drives our the peer influence.

\paragraph{Including Those Who Were Not in the Armies} 

As discussed above, peer composition is not systematically correlated with one's choice of participating in the war or not. Thus, we have focused on joining the Union versus the Confederacy for simplicity. If we further employ a multinomial logit model to consider whether a cadet joined the Union, joined the Confederacy or did not participate in the war, we again find that peer composition is not predictive for whether a cadet participated in the war, while our main finding of peer influence on joining the Union vs Confederacy holds. These results are presented in Supplemental Appendix Table \ref{Atab_mlogit}. According to these estimates on the log of the odds ratios, a one-standard deviation increase in Free-State peers increased probability of joining the Union by approximately 7 percentage points, higher than our baseline estimates.

\paragraph{Considering Cash-Crop Employment}

\citet{white2024rebel} finds that a cadet's history of employment in cash-crop agriculture (``the graduate was recorded as having spent time as a “planter” (a plantation owner) or held any job related to the production, processing, sale, or export of cotton, indigo, rice, sugar, or tobacco'') is negatively associated with the likelihood of joining the Union.   That relationship that holds in our data as well (Supplemental Appendix Table \ref{tab_baseline_cashcrop}). Our primary interest, however, is in whether peer influence is moderated by this employment background.

Unlike the results based on slavery proxies, we find no clear interaction between peer influence and cash-crop employment history (Supplemental Appendix Table \ref{tab_baseline_cashcrop}, Columns (4) and (8)). This may be due in part to the relatively small proportion---only 4\%---of cadets from Slave States having such employment experience. 

\paragraph{Considering Appointment State}

We have seen that a cadet's home state serves as an important reference point for allegiance norms. We also examine the appointment state (which heavily overlaps with the home state) and calculate peer composition using the same method as in our main analysis. The analysis is robust to this alternative, as seen in Supplemental Appendix Table \ref{tab_appointment_state}, we obtain a similar comparable peer effect, as appointment state and home state were often the same.

\paragraph{Career Outcomes}

We observe the post-war outcomes of many who joined an army, allowing us to investigate how their decisions during this critical historical juncture influenced their ex post life results, such as military rank and survival probability. 
We do not presume that this had any influence on their allegiances, but it does trace some consequences of their decisions.

As indicated in Columns (1)-(2) of Supplemental Appendix Table \ref{tab_careeroutcome}, joining the Union correlates with a lower military rank in 1865 and a reduced probability of attaining a rank of General-level positions. This aligns with the fact that the Confederate army, was both smaller and had more generals per enlisted man, and starting from scratch had no existing officers, enabling quicker promotions and a higher likelihood of becoming a general for people with officer training.  On the other hand, joining the Union is associated with a lower risk of dying in the Civil War, reflecting the higher fatality rate on the Confederacy side, as reported in Column (3).

Motivated our earlier analysis, we employ peer exposure and home state as instrumental variables for joining the Union. The magnitudes of the IV estimate resemble the OLS estimates, as shown in Columns (4)-(6). As previously mentioned, cadets from states with a large slave population (more than one third) did not respond significantly to peer influence. Omitting those cadets deliver similar estimates, despite the reduction in sample size, as seen in Panel B.


Moreover, in Supplemental Appendix Table \ref{Atab_academic}, we investigate the correlation between academic rank and military rank in both the Union and Confederate armies. We find that West Point performance was more predictive of outcomes for the Union Army and less so for the Confederacy.

\section{Discussion}

Our study highlights the significant role of peer exposure in shaping West Point cadets' allegiance choices during the American Civil War, especially for those from Slave States who faced a tension between nationalism and sectionalism among other tensions. We find that the strength of peer influence was modulated by the economic interests of cadets' home states: in regions where the slave economy was large, the influence of peers was less pronounced. 
This suggests that when economic stakes were particularly high and slavery more prevalent, cadets were less susceptible to peer-driven decisions. This finding is also consistent with the fact that states with larger slave shares were the first to secede and the level of debate in states was at least partly reflective of slave shares \citep{crofts2014reluctant}. 

Although our analysis is a case study, it is a useful proof of concept.
The decisions that West Point cadets made during the U.S. Civil War parallel those that arise in many divided societies, where individuals must choose between between polarized factions during critical moments.  
We have shown that in such a context, peer interactions can still play a decisive and significant role. 

Our study also provides a possibility for identifying such peer effects in other high-stakes and polarized settings.
This sort of peer influence is challenging to quantify and identify causally through anecdotal evidence alone.  It is also not enough to identify correlation in people's behaviors, given that friendships are endogenous and people are subject to common (potentially unobserved) factors \citep{angrist2014perils}. West Point's environment---characterized by structured, close-knit interactions among cadets from diverse geographic backgrounds---provided  us an opportunity to analyze the influence of quasi-random peer exposure, and its interplay with background allegiances. 
By using variation in a cadet's peers' state-of-origin for identification, we have something that is exogenous to other key things that influenced a given cadet's decision.  Similar techniques can be used in other high-stakes contexts.

We provide a simple model that can indicate why
the peer influence only affected cadets from states with slaves, and primarily among those cadets who did not own slaves nor lived in regions heavily dependent upon slaves.  
Our evidence on the strength of within-cohort influence, the relevance of continuous military experience, and how joint participation in the Mexican-American War amplifies peer influence suggests that shared-experiences mattered. 

We are still left with the fact that peer effects could have operated in (at least) three ways: (1) by fostering communication and persuasion, whereby cadets from Slave States may have been convinced to support the Union cause through exposure to pro-Union peers, (2) by building friendship bonds that influenced cadets' decisions to fight alongside their peers, or (3) by shifting perceptions of the likely outcome of the war, where cadets from Free States could have led their peers to believe that the Union was more likely to prevail. Our data do not allow us to disentangle these, or other mechanisms, and all could have operated together (which would be consistent with some evidence from personal letters and diaries \citep{mcpherson1997cause}). Further investigation of peer influence on allegiances in modern contexts, where additional data are more readily available, could offer additional insights into how such mechanisms operate.

\section*{Materials and Methods}

\paragraph{Data}
We construct a dataset of 1,638 West Point cadets who graduated between 1820 and 1860. The data are based on a manual digitization of biographical records from the \textit{Biographical Register of the Officers and Graduates of the United States Military Academy} \citep{cullum1891biographical}, supplemented with \textit{Rebels from West Point} \citep{patterson2002rebels}, \textit{Southern Historical Society Papers} \citep{herndon1876southern}, and \textit{Confederate Military History} \citep{evans1899confederate}. Civil War allegiance is coded according to whether a cadet served with Union or Confederate forces.

We augment these records with individual- and family-level characteristics, including slave ownership, parental origins, and spousal origins, by manually linking cadets to genealogical databases (Ancestry.com and FamilySearch.com) and to the 1860 U.S. Census, including the Slave Schedules. States are classified as Free or Slave based on whether slavery was legal in 1860; in our baseline analysis, this corresponds to a slave population share exceeding 1\%. Full details of data construction are provided in Supplemental Appendix \ref{Dataconstruct}.

\paragraph{Empirical Strategy}

Our identification strategy exploits plausibly quasi-random variation in cohort composition generated by the decentralized congressional nomination process and qualification examinations. These institutional features induce substantial year-to-year fluctuations in the share of peers from Free States that are orthogonal to individual cadet characteristics.

We estimate the impact of peer composition on allegiance decisions using linear probability models with controls for individual characteristics, state fixed effects, and cohort trends. Standard errors are obtained via bootstrapping, and we report additional specifications clustering at the cohort level. Robustness checks and additional analyses are presented in Supplemental Appendix \ref{a_sec_results}.

Data and Code Availability:  data and code have been deposited in Github (\url{https://github.com/GuoYuchen53/West-Point}).

        \setcitestyle{numbers} 
        \bibliographystyle{aeaown}
	\bibliography{reference}
\clearpage	

\pagenumbering{arabic}
\renewcommand*{\thepage}{A-\arabic{page}}
\setcounter{table}{0} 
\renewcommand\thetable{S\arabic{table}} 

\setcounter{figure}{0} 
\renewcommand\thefigure{S\arabic{figure}} 
\captionsetup[table]{aboveskip=0pt}
\captionsetup[table]{belowskip=0pt}

\appendix
\addcontentsline{toc}{section}{Supplemental Appendix} 
\begin{center}
    \Large{\textbf{Supplemental Appendix for ``Comrades and Cause: Peer Influence on West Point Cadets' Civil War Allegiances,'' by Guo, Jackson, Jia}} %
\end{center} 
\parttoc 

\section{Data Construction and Description}\label{Dataconstruct}

\subsection{Data Construction} \label{A_data}

\begin{figure}[ptb]
	\begin{minipage}{\linewidth}
		\vspace{3pt}
		\centerline{\includegraphics[width=\textwidth]{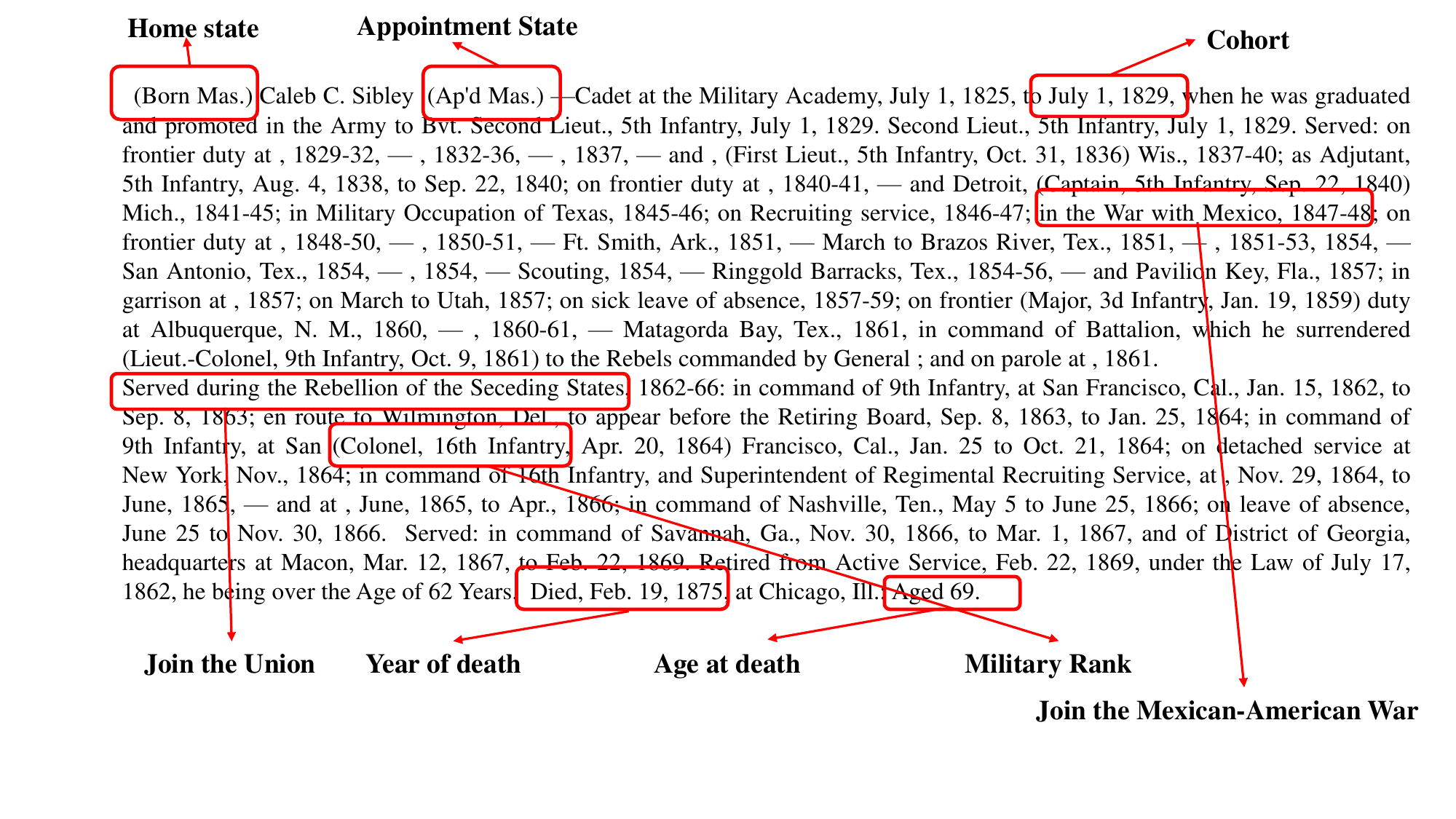}}
		\footnotesize\centerline{A: Example of coding the the Union}
	\end{minipage}
 
	\begin{minipage}{\linewidth}
		\vspace{3pt}
		\centerline{\includegraphics[width=\textwidth]{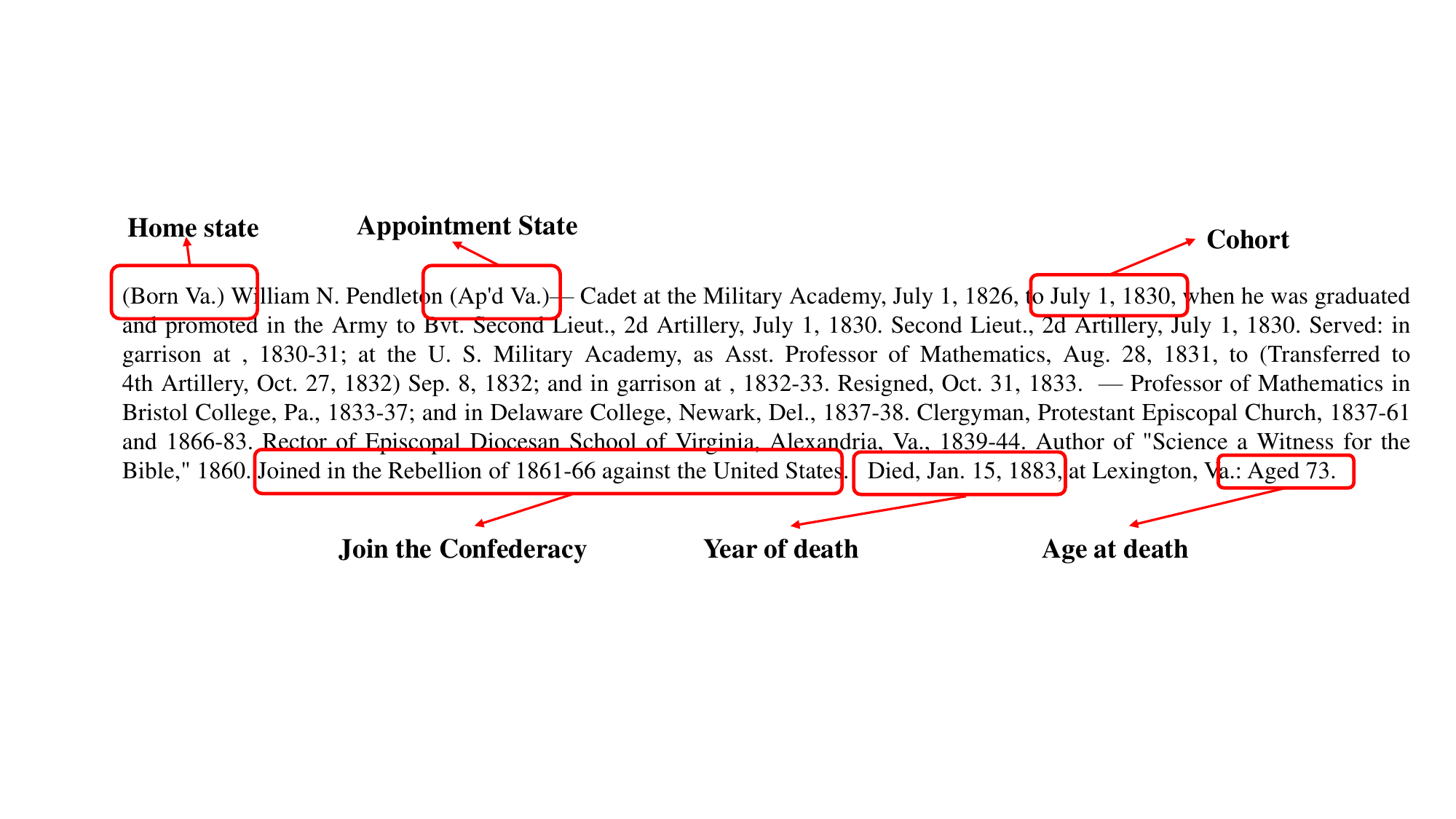}}
	 
		\footnotesize\centerline{B: Example of coding the Confederacy}
	\end{minipage}
 
	\caption{Text examples} \label{fig_examples_coding}
 \vspace{0.2cm}
\begin{spacing}{0.75}
{\footnotesize {\scriptsize \textit{Note.} This figure presents text examples from \citet{cullum1891biographical}. We use these texts as our main data source and complement the data with information from additional sources}}
\end{spacing}
\end{figure}

We construct a dataset of West Point cadets from 1820 to 1860. We manually collect and organize information on graduates, including their Union or Confederate affiliation during the Civil War, class ranking at graduation, place of birth, year of birth, and military rank in 1865, from the \textit{Biographical Register of the Officers and Graduates of the United States Military Academy} and other sources. We first present two examples of coding from main source in Figure \ref{fig_examples_coding}. We further supplement the dataset with additional individual- and family-level information from \textit{Ancestry.com} and \textit{FamilyTree.com} (and linked US Census data), including slaveholding status, parental information, spousal information, etc.

\paragraph{Civil War Affiliation} 
We determine military allegiance based on whether the individual served with the Union or the Confederacy during the Civil War. This is exemplified by statements such as ``He took part in the Rebellion of 1861-66 against the United States'' or ``He served in the Rebellion of the Seceding States.'' 

For those whose biographies do not clearly specify a side, we deduce their affiliation from the forces they served with during the war, including volunteer armies and local militias. For instance, service in the New York State Military during the Civil War categorizes them as Union members. 

We cross reference our data with additional sources including \textit{Find a Grave}, Wikipedia, \textit{Rebels from West Point}, and \textit{Southern Historical Society Papers}.

\paragraph{Class Rank at Graduation}

Since the number of cadets graduating in each West Point cohort varied, direct comparisons of class ranks across different cohorts are not meaningful. To ensure comparability, we classify cadets into percentiles based on their graduation ranking. The top 10\% were assigned a score of 100, the bottom 10\% a score of 10, and so on.

\paragraph{Year of Birth} 
In a few cases where an individual's year of birth was unavailable in the \textit{Biographical Register} , we supplement this information by using details from \textit{Find a Grave} and \textit{FamilySearch}.

\paragraph{Birth Place} 
The birthplace information reported in the \textit{Biographical Register of the Officers and Graduates of the United States Military Academy} is occasionally incomplete or imprecise, particularly with respect to county-level identifiers. To address these limitations, we supplement the original records using genealogical sources, including \textit{Find a Grave} and \textit{FamilySearch}, which provide more detailed and consistent location information. These sources allow us to recover missing county-level data and to correct ambiguous or inaccurate birthplace entries, thereby improving the geographic precision of our measures.

\paragraph{Military Rank in 1865}

For cadets who joined the Union, we determine their official military rank in 1865 from personal biographies. For cadets who joined the Confederacy, we refer to sources including \textit{Rebels from West Point}, \textit{Southern Historical Society Papers}, \textit{Confederate Military History}, and Wikipedia, for military rank descriptions.  

If the personal information did not include a specific military rank in 1865, we approximated it based on the individual's military career trajectory. For example, if an individual's career record states: \{'1846: Second Lieutenant', '1864: Major', '1870: Colonel'\}, we assign the rank held closest to 1865, but no later than 1865. In this case, the individual's rank in 1865 would be Major. Since every change in a person's military rank is recorded, if there is no mention of a change, we assume the rank remained the same. We focus on formal military ranks rather than honorary ranks and volunteer military ranks. Military ranks are grouped into 11 levels, with General ranked as 11 and Third Lieutenant ranked as 1.

\paragraph{Participation in the Mexican-American War}

The information is constructed from the \textit{Biographical Register of the Officers and Graduates of the United States Military Academy}. We record individuals who directly participated in the Mexican-American War, including those engaged in combat or other military operations. 

\paragraph{Genealogical and Census Data Linkage}

We supplemented the West Point biographical records with individual- and family-level information via manual linkage to genealogical (\textit{FamilySearch}) and US Census databases (\textit{Ancestry}). Using cadets' names and birth/death dates, we first searched \textit{FamilySearch} to identify the correct individual, recorded the unique page identifier, and collected birthplace, birthdate, residence around 1860, basic marital information and parental information. We then searched \textit{Ancestry} for the 1860 Census and confirmed matches based on consistency in name, age, birthplace, and residence, extracting occupation, residence, and reported real and personal estate values. Finally, we consulted the 1860 Slave Schedules on \textit{Ancestry}, using name and residence consistency to identify slaveholding status. Using cadets' county of residence, we further linked each individual to county-level characteristics in 1860, including the number of churches\footnote{Inter-university Consortium for Political and Social Research. United States Historical Election Returns, 1824-1968.} and the voting electoral returns\footnote{Haines, Michael R., and Inter-university Consortium for Political and Social Research. Historical, Demographic, Economic, and Social Data: The United States, 1790-2002.}.


\clearpage
\subsection{Classification of States} \label{A_states}
\begin{table}[H]
		
   \caption{ \centering Classification of states} \label{Atab_states}
		
    \begin{center}
			
			\footnotesize \begin{tabular}{l*{1}{cccccc}}
\hline\hline
	 & Slave share (\%) & 1\% threshold & 5\% threshold & 10\% threshold & 4 groups \\
\hline
	South Carolina  & 57.2 & Slave-State  & $>$5\%  & $>$10\% & Heavy-Slave State  \\
	Mississippi  & 55.2 & Slave-State  & $>$5\%  & $>$10\% & Heavy-Slave State  \\
	Louisiana  & 46.9 & Slave-State  & $>$5\%  & $>$10\% & Heavy-Slave State  \\
	Alabama  & 45.1 & Slave-State  & $>$5\%  & $>$10\% & Heavy-Slave State  \\
	Florida  & 44.0 & Slave-State  & $>$5\%  & $>$10\% & Heavy-Slave State  \\
	Georgia  & 43.7 & Slave-State  & $>$5\%  & $>$10\% & Heavy-Slave State  \\
	North Carolina  & 33.7 & Slave-State  & $>$5\%  & $>$10\% & Heavy-Slave State  \\
	Virginia  & 30.7 & Slave-State  & $>$5\%  & $>$10\% & Mid-Slave State  \\
	Texas  & 30.2 & Slave-State  & $>$5\%  & $>$10\% & Mid-Slave State  \\
	Arkansas  & 25.5 & Slave-State  & $>$5\%  & $>$10\% & Mid-Slave State  \\
	Tennessee  & 24.8 & Slave-State  & $>$5\%  & $>$10\% & Mid-Slave State  \\
	Kentucky  & 19.5 & Slave-State  & $>$5\%  & $>$10\% & Border-State  \\
	Maryland  & 12.7 & Slave-State  & $>$5\%  & $>$10\% & Border-State  \\
	Missouri  & 9.7 & Slave-State  & $>$5\%  & $<$10\% & Border-State  \\
	District of Columbia  & 4.4 & Slave-State &$<$5\%  & $<$10\%  &Border-State  \\
	Delaware  & 1.6 & Slave-State  & $<$5\%  & $<$10\% & Border-State  \\
        New Jersey  & 0.01 & Free-State  & $<$5\%  & $<$10\% & Free-State  \\
	New York  & 0.0 & Free-State  & $<$5\%  & $<$10\% & Free-State  \\
	Pennsylvania  & 0.0 & Free-State  & $<$5\%  & $<$10\% & Free-State  \\
	Ohio  & 0.0 & Free-State   & $<$5\%  & $<$10\%  & Free-State  \\
	Illinois  & 0.0 & Free-State  & $<$5\%  & $<$10\%  & Free-State \\
	Indiana  & 0.0 & Free-State  & $<$5\%  & $<$10\% & Free-State  \\
	Massachusetts  & 0.0 & Free-State  & $<$5\%  & $<$10\% & Free-State  \\
	Wisconsin  & 0.0 & Free-State  & $<$5\%  & $<$10\% & Free-State  \\
	Michigan  & 0.0 & Free-State  & $<$5\%  & $<$10\% & Free-State  \\
	Iowa  & 0.0 & Free-State  & $<$5\%  & $<$10\% & Free-State  \\
	Maine  & 0.0 & Free-State  & $<$5\%  & $<$10\%  & Free-State \\
	Connecticut  & 0.0 & Free-State  & $<$5\%  & $<$10\% & Free-State  \\
	California  & 0.0 & Free-State  & $<$5\%  & $<$10\% & Free-State  \\
	New Hampshire  & 0.0 & Free-State  & $<$5\%  & $<$10\% & Free-State  \\
	Vermont  & 0.0 & Free-State  & $<$5\%  & $<$10\%  & Free-State \\
	Rhode Island  & 0.0 & Free-State  & $<$5\%  & $<$10\% & Free-State  \\
	Minnesota  & 0.0 & Free-State  & $<$5\%  & $<$10\% & Free-State \\
	Oregon  & 0.0 & Free-State  & $<$5\%  & $<$10\% & Free-State  \\
			\hline\hline
		\end{tabular}

	\end{center}
 \vspace{0.2cm}
\begin{spacing}{0.75}
{\footnotesize {\scriptsize \textit{Note.} This table presents different ways we employ to classify home states of West Point cadets.}}
\end{spacing}
\end{table}


\clearpage
\subsection{Summary Statistics} \label{A_summary}
 
\begin{table}[H]
	\renewcommand{\arraystretch}{0.93}	
   \caption{ \centering Summary statistics} \label{Atab_summary}
		
    \begin{center}
			
	\footnotesize		\begin{tabular}{l*{1}{ccccc}}
\hline\hline
		&  Mean&  Std.\ Dev.& Min.& Max.& Obs.\\
\hline
Panel A: Heavy-Slave States (slave share $>$ 33\%) \\
\hline
		Joining the Union (War Participants)  & 0.107 & 0.310 & 0 & 1 & 122 \\
  	    Joining the Union (All)  & 0.062&  0.242& 0 & 1 & 210 \\
        Joining the War   & 0.581&  0.495& 0 & 1 & 210 \\

       \\
		Class Rank & 53.810 & 30.068 & 10 & 100 & 210 \\
		Age in 1860 & 40.530 & 11.315 & 23 & 65 & 210\\
            Slave Pop. Share (state level) & 45.370 & 9.467 & 33.4 & 57.2 & 210\\
            Cohort & 1841.290 & 11.716 & 1820 & 1860 & 210\\
            Join Mex.-Am. War & 0.276 & 0.448 & 0 & 1 & 210\\
            
            \\
        Cadet Slave Ownership & 0.352 & 0.480 & 0 & 1 & 122\\   
		Slave Pop. Share (County-level) & 49.415 & 16.553 & 13.191 & 84.999 & 108\\
        Free-State Father & 0.204 & 0.405 & 0 & 1 & 103\\   
        Free-State Mother & 0.080 & 0.273 & 0 & 1 & 100\\
        Free-State Wife   & 0.172 & 0.379 & 0 & 1 & 122\\
\hline
Panel B: Border and Mid-Slave States ( 1\%$<$slave share $\leq$ 33\%) \\
\hline
		Joining the Union (War Participants)  & 0.477 & 0.500 & 0 & 1 & 266 \\
  	    Joining the Union (All)  & 0.276&  0.448& 0 & 1 & 460 \\
        Joining the War  & 0.578&  0.494& 0 & 1 & 460 \\

       \\
		Class Rank & 54.326 & 28.288 & 10 & 100 & 460 \\
		Age in 1860 & 41.841 & 10.461 & 22 & 62 & 460 \\
            Slave Pop. Share & 20.739 & 9.379 & 1.6 & 30.700 & 460\\
            Cohort & 1839.761 & 10.890 & 1820 & 1860 & 460\\
            Join Mex.-Am. War & 0.317 & 0.466 & 0 & 1 & 460\\
            
            \\
        Slave Ownership & 0.207 & 0.406 & 0 & 1 & 266\\   
		Slave Pop. Share (County-level) & 24.269 & 18.718 & 0.446 & 71.269 & 244\\
        Free-State Father & 0.132 & 0.340 & 0 & 1 & 234\\   
        Free-State Mother & 0.137 & 0.345 & 0 & 1 & 233\\
        Free-State Wife   & 0.297 & 0.458 & 0 & 1 & 266\\			
\hline
Panel C: Free States (slave share $<$ 1\%) \\
\hline
		Joining the Union (War Participants)  & 0.919 & 0.274 & 0 & 1 & 540 \\
  	    Joining the Union (All)  & 0.512&  0.500& 0 & 1 & 968 \\
        Joining the War  & 0.558&  0.497& 0 & 1 & 968 \\

       \\
		Class Rank & 57.066 & 28.663 & 10 & 100 & 968 \\
		Age in 1860 & 41.535 & 10.899 & 21 & 63 & 968 \\
            Slave Pop. Share & 0.000 & 0.002 & 0 & 0.010 & 968\\
            Cohort & 1840.346 & 11.337 & 1820 & 1860 & 968\\
            Join Mex.-Am. War & 0.305 & 0.461 & 0 & 1 & 968\\
            
            \\
        Cadet Slave Ownership & 0 & 0 & 0 & 0 & 540\\   
		Slave Pop. Share (County-level) & 0.000 & 0.000 & 0 & 0.012 & 433\\
        Free-State Father & 0.868 & 0.338 & 0 & 1 & 471\\   
        Free-State Mother & 0.890 & 0.314 & 0 & 1 & 462\\
        Free-State Wife   & 0.467 & 0.499 & 0 & 1 & 540\\	
            
   \hline\hline
		\end{tabular}

	\end{center}
{\footnotesize {\scriptsize \textit{Note.} This table presents summary statistics for the key variables used in our analysis. \textit{Slave Pop. Share (county level)} and \textit{Slave Pop. Share (state level)} are expressed in percentages. Variables \textit{Free-State Father}, \textit{Free-State Mother}, \textit{Free-State Wife}, and \textit{Cadet Slave Ownership} are collected only for individuals who participated in the Civil War.}}
\end{table}


\clearpage
 
\begin{table}[H]
		
   \caption{ \centering Summary statistics: Cohort-level} \label{Atab_cohort_summary}
		
    \begin{center}
			
	\footnotesize		\begin{tabular}{l*{1}{ccccc}}
\hline\hline
		&  Mean&  Std.\ Dev.& Min.& Max.& Obs.\\
\hline
Panel A: Heavy-Slave States (slave share $>$ 33\%) \\
\hline
		Number of Graduates      &  5.122 & 2.532 & 0     & 11    & 41 \\
Fraction of Graduates    &  0.127 & 0.057 & 0     & 0.263 & 41 \\
\hline
Panel B: Border and Mid-Slave States ( 1\%$<$slave share $\leq$ 33\%) \\
\hline
		Number of Graduates      & 11.220 & 3.863 & 2     & 19    & 41 \\
Fraction of Graduates    &  0.280 & 0.080 & 0.059 & 0.480 & 41 \\		
\hline
Panel C: Free States (slave share $<$ 1\%) \\
\hline
		Number of Graduates      & 23.610 & 5.572 & 12    & 33    & 41 \\
Fraction of Graduates    &  0.593 & 0.085 & 0.462 & 0.824 & 41 \\
            
   \hline\hline
		\end{tabular}

	\end{center}
 \vspace{0.2cm}
\begin{spacing}{0.75}
{\footnotesize {\scriptsize \textit{Note.} This table presents summary statistics at the  
cohort level.}}
\end{spacing}
\end{table}




\clearpage

\subsection{Correlations between Slave Share and Allegiances}


\begin{figure}[H]
	\begin{center}
	\begin{minipage}{0.5\linewidth}
		\vspace{3pt}
		\centerline{\includegraphics[width=\textwidth,trim=5 5 5 5,clip]{Figures/all_SlaveShareandJoinUnion.pdf}}
		\scriptsize\centerline{A: Probability of joining the Union}
	\end{minipage}
 
	\begin{minipage}{0.5\linewidth}
		\vspace{3pt}
		\centerline{\includegraphics[width=\textwidth,trim=5 5 5 5,clip]{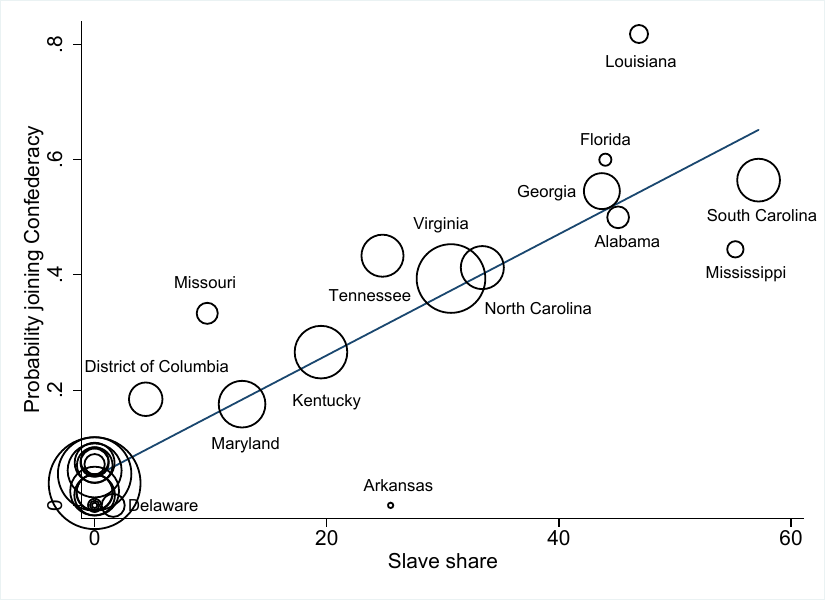}}
	 
		\scriptsize\centerline{B: Probability of joining the Confederacy}
	\end{minipage}
 
	\begin{minipage}{0.5\linewidth}
		\vspace{3pt}
		\centerline{\includegraphics[width=\textwidth,trim=5 5 5 5,clip]{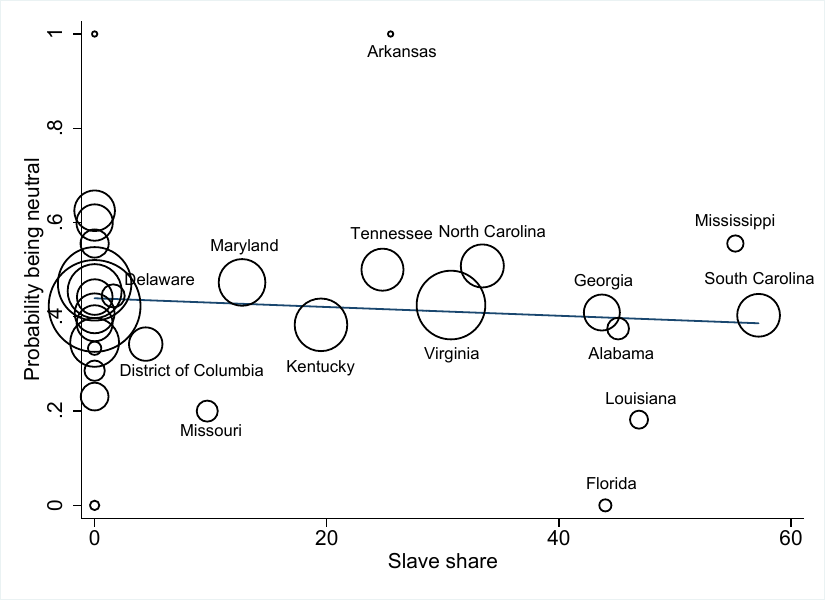}}
	 
		\scriptsize\centerline{C: Probability of being neutral}
	\end{minipage}
 
	\caption{Slave share and cadet's choices} \label{fig_byslave_Share_all}
 \end{center}
  \vspace{0.2cm}
\begin{spacing}{0.75}
{\footnotesize {\scriptsize \textit{Note.} The sample for this analysis consists of all cadets, including those who did not participate in the war. This figure illustrates the correlations between the proportion of slaves in a state and the likelihood of a cadet from that state opting to join the Union (A), the Confederacy (B), or abstain from participating (C).}}
\end{spacing}
\end{figure}

\clearpage
\subsection{Peers and Personal Characteristics}
\label{sec:identification}

\begin{table}[H]
		
   \caption{ \centering Correlations between peers and personal characteristics}\label{tab_validation}
		
    \begin{center}
			
	\footnotesize		
    \def\sym#1{\ifmmode^{#1}\else\(^{#1}\)\fi}
        \begin{tabular}{l*{8}{c}}
            \hline\hline
        Dependent var (sd)  
        &\multicolumn{8}{c}{Share Free-State peers:1820-1860} 
        \\\cmidrule(lr){2-9}
        &\multicolumn{4}{c}{War Participants}
        &\multicolumn{4}{c}{All}
        \\\cmidrule(lr){2-5}\cmidrule(lr){6-9}
            &\multicolumn{2}{c}{ Slave States }
            &\multicolumn{2}{c}{ Free States }
            &\multicolumn{2}{c}{ Slave States }
            &\multicolumn{2}{c}{ Free States }
            \\\cmidrule(lr){2-3}\cmidrule(lr){4-5}
            \cmidrule(lr){6-7}\cmidrule(lr){8-9}
            &\multicolumn{1}{c}{(1)}
            &\multicolumn{1}{c}{(2)} 
            &\multicolumn{1}{c}{(3)} 
            &\multicolumn{1}{c}{(4)}
            &\multicolumn{1}{c}{(5)} 
            &\multicolumn{1}{c}{(6)} 
            &\multicolumn{1}{c}{(7)} 
            &\multicolumn{1}{c}{(8)}\\
            \hline
            Age in 1860   &   -0.004         &    0.012         &   -0.004         &   -0.003         &   -0.007         &    0.012         &    0.005         &   -0.003         \\
                &  (0.031)         &  (0.037)         &  (0.027)         &  (0.028)         &  (0.026)         &  (0.037)         &  (0.019)         &  (0.028)         \\
            Class Rank     &   -0.002         &   -0.002         &    0.001         &    0.001         &   -0.001         &   -0.002         &    0.000         &    0.001         \\
                &  (0.002)         &  (0.002)         &  (0.001)         &  (0.002)         &  (0.001)         &  (0.002)         &  (0.001)         &  (0.002)         \\
            Slave Pop. Share (sd)&   -0.107\sym{**} &   -0.069         &   -0.001         &    0.011         &   -0.030         &   -0.069         &   -0.035         &    0.011         \\
                &  (0.049)         &  (0.063)         &  (0.042)         &  (0.040)         &  (0.038)         &  (0.063)         &  (0.030)         &  (0.040)         \\
            
            Cohort (t)   &   -0.002         &    0.017         &    0.007         &    0.009         &   -0.009         &    0.017         &    0.010         &    0.009         \\
                &  (0.030)         &  (0.036)         &  (0.027)         &  (0.028)         &  (0.025)         &  (0.036)         &  (0.018)         &  (0.028)         \\
Free-State Father &                  &   -0.005         &                  &    0.013         &                  &   -0.005         &                  &    0.013         \\
                &                  &  (0.186)         &                  &  (0.156)         &                  &  (0.186)         &                  &  (0.156)         \\
Free-State Mother &                  &    0.347         &                  &   -0.124         &                  &    0.347         &                  &   -0.124         \\
                &                  &  (0.216)         &                  &  (0.182)         &                  &  (0.216)         &                  &  (0.182)         \\
Free-State Wife &                  &   -0.073         &                  &    0.033         &                  &   -0.073         &                  &    0.033         \\
                &                  &  (0.141)         &                  &  (0.101)         &                  &  (0.141)         &                  &  (0.101)         \\
            \hline
 Joint test $p$-value&    0.190    &   0.336        &       0.130      &   0.370    &      0.739       &  0.336       &   0.212         &   0.370        \\
            \hline
            Observations    &      388         &      324         &      540         &      456         &      670         &      324         &      968         &      456         \\
            R-squared       &    0.015         &    0.027         &    0.013         &    0.017         &    0.003         &    0.027         &    0.006         &    0.017         \\
            \hline\hline
        \end{tabular}

	\end{center}
\par
\vspace{0.2cm}
\begin{spacing}{0.75}
{\footnotesize {\scriptsize \textit{Note.} This table presents the correlations between peer composition and individual traits. The joint test on overall correlations are statistically undistinguished from zero. The standard errors presented in the parentheses are obtained through bootstrapping with 400 resampling iterations. ***$p<0.01$, **$p<0.05$, *$p<0.1$.}}
\end{spacing}

\end{table}

\clearpage

\subsection{Alternative Thresholds}  \label{A_thresholds2}
\begin{table}[h]
		
   \caption{ \centering Peer composition and allegiance choice: Different thresholds}\label{Atab_thresholds2}

    \begin{center}

    \def\sym#1{\ifmmode^{#1}\else\(^{#1}\)\fi}
    \resizebox{\textwidth}{!}{
        \begin{tabular}{l*{8}{c}}
            \hline\hline
            Dependent var
            &\multicolumn{8}{c}{Join the Union: War Participants}
            \\\cmidrule(lr){2-9}
            &\multicolumn{4}{c}{Thresholds: 5\%} 
            &\multicolumn{4}{c}{Thresholds: 10\%}
            \\\cmidrule(lr){2-5}\cmidrule(lr){6-9}
            &\multicolumn{2}{c}{Slave States} 
            &\multicolumn{2}{c}{Free States}
            &\multicolumn{2}{c}{Slave States} 
            &\multicolumn{2}{c}{Free States}
            \\\cmidrule(lr){2-3}\cmidrule(lr){4-5}\cmidrule(lr){6-7}\cmidrule(lr){8-9}
            &\multicolumn{1}{c}{(1)} 
            &\multicolumn{1}{c}{(2)} 
            &\multicolumn{1}{c}{(3)} 
            &\multicolumn{1}{c}{(4)}
            &\multicolumn{1}{c}{(5)} 
            &\multicolumn{1}{c}{(6)} 
            &\multicolumn{1}{c}{(7)} 
            &\multicolumn{1}{c}{(8)}\\
            \hline
            Share Free-State Peers (sd)   &    0.061\sym{***}&    0.065\sym{***}&   -0.008         &   -0.011        &    0.064\sym{***}&    0.064\sym{***}&   -0.007         &   -0.009       \\
                &  (0.022)         &  (0.022)         &  (0.012)         &  (0.012)       &  (0.024)         &  (0.024)         &  (0.013)         &  (0.013)          \\
            Age in 1860    &   -0.008         &   -0.003         &    0.004         &    0.003         &   -0.014         &   -0.009         &    0.006         &    0.005         \\
                &  (0.015)         &  (0.016)         &  (0.007)         &  (0.008)         &  (0.016)         &  (0.016)         &  (0.007)         &  (0.007)         \\
            Class Rank    &    0.001         &    0.001         &   -0.001\sym{**} &   -0.001\sym{**} &    0.001         &    0.001         &   -0.001\sym{**} &   -0.001\sym{**} \\
                &  (0.001)         &  (0.001)         &  (0.000)         &  (0.000)         &  (0.001)         &  (0.001)         &  (0.000)         &  (0.000)         \\
            Slave Pop. Share (sd) &   -0.182\sym{***}&                  &   -0.033\sym{*}  &                  &   -0.177\sym{***}&                  &   -0.060\sym{***}&                  \\
                &  (0.021)         &                  &  (0.018)         &                  &  (0.021)         &                  &  (0.020)         &                  \\
            Cohort     &   -0.009         &   -0.004         &    0.007         &    0.006         &   -0.015         &   -0.009         &    0.008         &    0.007         \\
                &  (0.015)         &  (0.015)         &  (0.007)         &  (0.007)         &  (0.015)         &  (0.016)         &  (0.007)         &  (0.007)         \\
            State FEs &N&Y&N&Y&N&Y&N&Y\\
        \hline
        Dependent var. mean&0.317&0.317&0.911&0.911&0.308&0.308&0.905&0.905\\
        Observations    &      353         &      353         &      575         &      575         &      341         &      341         &      587         &      587         \\
R-squared       &    0.179         &    0.238         &    0.034         &    0.052         &    0.178         &    0.238         &    0.055         &    0.072         \\
        \hline\hline
        \end{tabular}}

	\end{center}
\par
\vspace{0.2cm}
\begin{spacing}{0.75}
{\footnotesize {\scriptsize \textit{Note.}  The standard errors reported in the parentheses are derived from bootstrapping with 400 resampling iterations. ***$p<0.01$, **$p<0.05$, *$p<0.1$.}}
\end{spacing}

\end{table}

\clearpage
\section{Additional Results} \label{a_sec_results}

\subsection{Cohort-Clustered and Bootstrap Cohort-Clustered SE} 

\begin{table}[h]
	
   \caption{ \centering Peer composition and allegiance choice: Variations on standard errors}\label{tab_different_SE}
		
    \begin{center}
			
			\footnotesize     \def\sym#1{\ifmmode^{#1}\else\(^{#1}\)\fi}
        \begin{tabular}{l*{8}{c}}
            \hline\hline
            Dependent var
            &\multicolumn{8}{c}{Join the Union: War Participants}
            \\\cmidrule(lr){2-7}
            &\multicolumn{4}{c}{Slave States} 
            &\multicolumn{4}{c}{Free States}
            \\\cmidrule(lr){2-5}\cmidrule(lr){6-9}
            &\multicolumn{1}{c}{(1)} 
            &\multicolumn{1}{c}{(2)} 
            &\multicolumn{1}{c}{(3)} 
            &\multicolumn{1}{c}{(4)}
            &\multicolumn{1}{c}{(5)} 
            &\multicolumn{1}{c}{(6)}
            &\multicolumn{1}{c}{(7)} 
            &\multicolumn{1}{c}{(8)}\\
\hline
A. Cohort level bootstrap \\
\hline
            Share Free-State Peers (sd) &    0.080\sym{***}&    0.058\sym{***}&    0.054\sym{***} &    0.056\sym{***} &   -0.003         &   -0.005         &   -0.006    &  -0.005       \\
                &  (0.020)         &  (0.014)         &  (0.014)   &  (0.017)       &  (0.015)         &  (0.015)         &  (0.015)   &  (0.015)       \\
            Controls &N&Y&Y&Y&N&Y&Y&Y\\
            State FEs &N&N&Y&Y&N&N&Y&Y\\
        \hline
        Dependent var. mean &0.361&0.361&0.361&0.355&0.919&0.919&0.919&0.930\\
ICC &0.000&0.000&0.000&0.000&0.028&0.016&0.015&0.025\\        
Observations    &      388         &      388         &      388         &      324         &      540         &      540         &      540         &      456         \\
R-squared       &    0.028         &    0.234         &    0.287         &    0.363         &    0.000         &    0.017         &    0.030         &    0.067         \\
\hline\hline
\hline
B. Cohort level cluster  \\
\hline
            Share Free-State Peers (sd) &    0.080\sym{***}&    0.058\sym{***}&    0.054\sym{***} &    0.056\sym{***}&   -0.003         &   -0.005         &   -0.006    &   -0.005      \\
                &  (0.020)         &  (0.013)         &  (0.019)  &  (0.020)       &  (0.012)         &  (0.012)         &  (0.013)    &  (0.012)      \\
            Controls &N&Y&Y&Y&N&Y&Y&Y\\
            State FEs &N&N&Y&Y&N&N&Y&Y\\
        \hline
        Dependent var. mean &0.361&0.361&0.361&0.355&0.919&0.919&0.919&0.930\\
ICC &0.000&0.000&0.000&0.000&0.028&0.016&0.015&0.025\\        
Observations    &      388         &      388         &      388         &      324         &      540         &      540         &      540         &      456         \\
R-squared       &    0.028         &    0.234         &    0.287         &    0.363         &    0.000         &    0.017         &    0.030         &    0.067         \\
        \hline\hline
        \end{tabular}

	\end{center}
\par
\vspace{0.2cm}
\begin{spacing}{0.75}
{\footnotesize {\scriptsize \textit{Note.} Panel A reports the re-estimation of the baseline results using robust standard errors clustered at the cohort level.
Panel B reports the corresponding re-estimates using bootstrapped standard errors clustered at the cohort level with 400 resampling iterations. 
(This table presents the impact of the fraction of peers from Free States in a cadet's cohort on that cadet's decision to join the Union. The control variables consist of \textit{Age in 1860}, \textit{Class Rank}, state level \textit{Slave Population Share (sd)}, and \textit{Cohort}. In addition, columns (4) and (8) further control for cadet slave ownership as well as family background characteristics, including whether the cadet’s father, mother, or wife was from a free state. Columns (1)-(4) focus on cadets from Slave States and Columns (5)-(8) on those from Free States. ***$p<0.01$, **$p<0.05$, *$p<0.1$.)}}
\end{spacing}

\end{table}

\clearpage

\subsection{Different Age Groups} 

\begin{table}[h]
	
   \caption{ \centering Peer composition and allegiance choice: Variations in age groups}\label{tab_different_age}
		
    \begin{center}
			
			\footnotesize     \def\sym#1{\ifmmode^{#1}\else\(^{#1}\)\fi}
        \begin{tabular}{l*{6}{c}}
            \hline\hline
            Dependent var
            &\multicolumn{6}{c}{Join the Union: War Participants}
            \\\cmidrule(lr){2-7}
            &\multicolumn{3}{c}{Slave States} 
            &\multicolumn{3}{c}{Free States}
            \\\cmidrule(lr){2-4}\cmidrule(lr){5-7}
            &\multicolumn{1}{c}{(1)} 
            &\multicolumn{1}{c}{(2)} 
            &\multicolumn{1}{c}{(3)} 
            &\multicolumn{1}{c}{(4)}
            &\multicolumn{1}{c}{(5)} 
            &\multicolumn{1}{c}{(6)}\\
\hline
A. $Age < 60$ \\
\hline
            Share Free-State Peers (sd)  &    0.074\sym{***}&    0.052\sym{**} &    0.048\sym{**} &   -0.002         &   -0.004         &   -0.005         \\
                &  (0.024)         &  (0.021)         &  (0.022)         &  (0.013)         &  (0.012)         &  (0.012)         \\
            Controls &N&Y&Y&N&Y&Y\\
            State FEs &N&N&Y&N&N&Y\\
        \hline
        Dependent var. mean &0.359&0.359&0.359&0.920&0.920&0.920\\
        Observations    &      384         &      384         &      384         &      535         &      535         &      535         \\
R-squared       &    0.024         &    0.236         &    0.288         &    0.000         &    0.014         &    0.027         \\
\hline\hline
\hline
B. $Age < 50$  \\
\hline
            Share Free-State Peers (sd)  &    0.080\sym{***}&    0.059\sym{***}&    0.053\sym{**} &   -0.005         &   -0.008         &   -0.009         \\
                &  (0.026)         &  (0.021)         &  (0.022)         &  (0.013)         &  (0.013)         &  (0.013)         \\
            Controls &N&Y&Y&N&Y&Y\\
            State FEs &N&N&Y&N&N&Y\\
        \hline
        Dependent var. mean &0.353&0.353&0.353&0.921&0.921&0.921\\
        Observations    &      334         &      334         &      334         &      471         &      471         &      471         \\
        R-squared       &    0.028         &    0.248         &    0.301         &    0.000         &    0.025         &    0.041         \\
        \hline\hline
        \end{tabular}

	\end{center}
\par
\vspace{0.2cm}
\begin{spacing}{0.75}
{\footnotesize {\scriptsize \textit{Note.} Panel A restricts the sample to cadets younger than 60 at the time of Civil War participation. Panel B further restricts the sample to those younger than 50. Columns (1)-(3) focus on cadets from Slave States and Columns (4)-(6) on those from Free States. ***$p<0.01$, **$p<0.05$, *$p<0.1$.)}}
\end{spacing}

\end{table}

\clearpage
\subsection{Home State Economic Proxies and War Allegiances}
\begin{table}[h]
		
   \caption{\centering Home state economic proxies and war allegiance}  \label{Atab_background_variables}

    \begin{center}
			
			\footnotesize     \def\sym#1{\ifmmode^{#1}\else\(^{#1}\)\fi}
    \resizebox{\textwidth}{!}{
        \begin{tabular}{l*{8}{c}}
            \hline\hline
            Dependent var
            &\multicolumn{8}{c}{Join the Union: War Participants}
            \\\cmidrule(lr){2-9}
            &\multicolumn{4}{c}{Slave States} 
            &\multicolumn{4}{c}{Free States}
            \\\cmidrule(lr){2-5}\cmidrule(lr){6-9}
            &\multicolumn{1}{c}{(1)} 
            &\multicolumn{1}{c}{(2)} 
            &\multicolumn{1}{c}{(3)} 
            &\multicolumn{1}{c}{(4)}
            &\multicolumn{1}{c}{(5)} 
            &\multicolumn{1}{c}{(6)}
            &\multicolumn{1}{c}{(7)} 
            &\multicolumn{1}{c}{(8)}\\
            \hline
            Share Free-State Peers (sd)   &    0.058\sym{***}&    0.058\sym{***}&    0.054\sym{***}&    0.053\sym{***}&   -0.005         &   -0.006         &   -0.005         &   -0.005         \\
                &  (0.020)         &  (0.020)         &  (0.020)         &  (0.020)         &  (0.013)         &  (0.013)         &  (0.013)         &  (0.013)         \\
            Age in 1860    &   -0.012         &   -0.012         &   -0.005         &   -0.005         &    0.006         &    0.006         &    0.006         &    0.006         \\
                &  (0.015)         &  (0.015)         &  (0.015)         &  (0.015)         &  (0.007)         &  (0.007)         &  (0.007)         &  (0.007)         \\
            Class Rank        &    0.001         &    0.001         &    0.000         &    0.000         &   -0.001\sym{**} &   -0.001\sym{*} &   -0.001\sym{*} &   -0.001\sym{*} \\
                &  (0.001)         &  (0.001)         &  (0.001)         &  (0.001)         &  (0.000)         &  (0.000)         &  (0.000)         &  (0.000)         \\
            Cohort         &   -0.013         &   -0.013         &   -0.005         &   -0.005         &    0.008         &    0.008         &    0.008         &    0.008         \\
                &  (0.015)         &  (0.015)         &  (0.015)         &  (0.015)         &  (0.007)         &  (0.007)         &  (0.007)         &  (0.007)         \\

            Slave Pop. Share (sd)&   -0.216\sym{***}&   -0.217\sym{***}&   -0.102\sym{**} &   -0.109\sym{***}&   -0.007         &   -0.011         &   -0.013         &   -0.012         \\
                &  (0.019)         &  (0.019)         &  (0.042)         &  (0.042)         &  (0.016)         &  (0.016)         &  (0.016)         &  (0.017)         \\
            ln Farmland Value per Capita&                  &    0.011         &    0.018         &    0.025         &                  &    0.052         &    0.072         &    0.069         \\
                &                  &  (0.048)         &  (0.048)         &  (0.048)         &                  &  (0.040)         &  (0.054)         &  (0.063)         \\
            ln Manufactured Product per Capita&                  &                  &    0.233\sym{***}&    0.168         &                  &                  &    0.019         &    0.022         \\
                &                  &                  &  (0.077)         &  (0.123)         &                  &                  &  (0.031)         &  (0.055)         \\
            Manufacturing Employment Share (sd)&                  &                  &                  &    0.035         &                  &                  &                  &   -0.002         \\
                &                  &                  &                  &  (0.056)         &                  &                  &                  &  (0.034)         \\
        \hline
        Dependent var. mean &0.361&0.361&0.361&0.361&0.919&0.919&0.919&0.919\\
        Observations    &      388         &      388         &      388         &      388         &      540         &      540         &      540         &      540         \\
        R-squared       &    0.234         &    0.234         &    0.252         &    0.253         &    0.017         &    0.020         &    0.021         &    0.021         \\
        \hline\hline
        \end{tabular}}

	\end{center}
\par
\vspace{0.2cm}
\begin{spacing}{0.75}
{\footnotesize {\scriptsize \textit{Note.} 
This table shows the association between war allegiances and several home state economic variables. The standard errors reported in the parentheses are derived from bootstrapping with 400 resampling iterations. ***$p<0.01$, **$p<0.05$, *$p<0.1$.}}
\end{spacing}

\end{table}

\clearpage

\subsection{Considering Religion and Voting} 

As a proxy for underlying religious attitudes, we follow Fogel \cite{fogel1994without} by classifying denominations into two broad groups: evangelical churches (Congregationalists, Methodists and Presbyterians) and liturgical churches (Catholics, Lutherans, Episcopalians) plus Quakers;   with the former group tending to be associated with pro-slavery constituencies, and the latter tending to be associated with anti-slavery constituencies in 1860, but with some being divided and/or not taking open stances. 
We use the number of churches of the given group per 1000 population as the measure.
Although this categorization of religion captures some basic tendencies to support or oppose slavery across the whole United States at the time, churches in slave states (other than the small set of Quakers) tended to support or be ambiguous on slavery regardless of category, which may be why we find no effects of religious categories in the table below. 

As expected, Republican vote share is positively associated with the probability of joining the Union. However, controlling for voting and its interaction with peer influence does not alter our main findings on slave population share or its interaction with peer effects, suggesting that slave share is a more powerful proxy for cadets’ political-economic background.

\begin{table}[h]
		
   \caption{ \centering Peer composition and allegiance choice: Religion and voting}\label{A_church_vote}

    \begin{center}
			
			 \def\sym#1{\ifmmode^{#1}\else\(^{#1}\)\fi}
 \resizebox{\textwidth}{!}{
        \begin{tabular}{l*{10}{c}}
            \hline\hline
            Dependent var
            &\multicolumn{10}{c}{Join the Union: War Participants}
            \\\cmidrule(lr){2-11}
            &\multicolumn{5}{c}{Slave States} 
            &\multicolumn{5}{c}{Border States}
            \\\cmidrule(lr){2-6}\cmidrule(lr){7-11}
            &\multicolumn{1}{c}{(1)} 
            &\multicolumn{1}{c}{(2)} 
            &\multicolumn{1}{c}{(3)} 
            &\multicolumn{1}{c}{(4)}
            &\multicolumn{1}{c}{(5)}
            &\multicolumn{1}{c}{(6)} 
            &\multicolumn{1}{c}{(7)} 
            &\multicolumn{1}{c}{(8)} 
            &\multicolumn{1}{c}{(9)}
            &\multicolumn{1}{c}{(10)}\\
            \hline
            Share Free-State Peers (sd)     &    0.053\sym{**} &    0.053\sym{**} &    0.054\sym{**} &    0.056\sym{**} &    0.054\sym{**} &    0.068\sym{***}&    0.067\sym{***}&    0.068\sym{***}&    0.073\sym{**} &    0.074\sym{**} \\
                &  (0.021)         &  (0.021)         &  (0.021)         &  (0.022)         &  (0.022)         &  (0.025)         &  (0.025)         &  (0.025)         &  (0.029)         &  (0.029)         \\
            Slave Share (County) $\times$  Share Free-State Peers (sd)      &   -0.036\sym{*}  &   -0.040\sym{**} &   -0.043\sym{**}   &   -0.047\sym{**} &   -0.041\sym{**}  &   -0.027         &   -0.025         &   -0.027          &   -0.046\sym{*}  &   -0.037         \\
                 &  (0.018)         &  (0.020)         &  (0.020)         &  (0.021)         &  (0.020)         &  (0.023)         &  (0.025)         &  (0.025)         &  (0.026)         &  (0.023)         \\
            Slave Share (County)   &   -0.166\sym{***}&   -0.170\sym{***}&   -0.165\sym{***}&   -0.142\sym{***}&   -0.161\sym{***}&   -0.187\sym{***}&   -0.187\sym{***}&   -0.184\sym{***}&   -0.163\sym{***}&   -0.183\sym{***}\\
                &  (0.033)         &  (0.033)         &  (0.034)         &  (0.034)         &  (0.034)         &  (0.036)         &  (0.036)         &  (0.036)         &  (0.036)         &  (0.034)         \\
                
            Evangelical $\times$  Share Free-State Peers (sd)&                  &   -0.012         &                  &                  &                  &                  &    0.011         &                  &                  &                  \\
                &                  &  (0.018)         &                  &                  &                  &                  &  (0.023)         &                  &                  &                  \\
            Evangelical     &                  &   -0.022  &                  &                  &                  &                  &    0.003         &                  &                  &                  \\
                &                  &  (0.026)         &                  &                  &                  &                  &  (0.033)         &                  &                  &                  \\
                
            Liturgical  $\times$  Share Free-State Peers (sd) &                  &                  &   -0.021         &                  &                  &                  &                  &   0.000         &                  &                  \\
                &                  &                  &  (0.024)         &                  &                  &                  &                  &  (0.026)         &                  &                  \\
            Liturgical + Quaker&                  &                  &    0.015         &                  &                  &                  &                  &    0.014         &                  &                  \\
                &                  &                  &  (0.036)         &                  &                  &                  &                  &  (0.045)         &                  &                  \\
         Republican $\times$  Share Free-State Peers (sd) &                  &                  &                  &   -0.013         &                  &                  &                  &                  &   -0.017         &                  \\
                &                  &                  &                  &  (0.018)         &                  &                  &                  &                  &  (0.022)         &                  \\
            Republican  &                  &                  &                  &    0.091\sym{**} &                  &                  &                  &                  &    0.090\sym{*}  &                  \\
                &                  &                  &                  &  (0.043)         &                  &                  &                  &                  &  (0.053)         &                  \\
            Southern Democrat $\times$  Share Free-State Peers (sd) &                  &                  &                  &                  &   -0.003         &                  &                  &                  &                  &    0.003         \\
                &                  &                  &                  &                  &  (0.023)         &                  &                  &                  &                  &  (0.029)         \\
            Southern Democrat &                  &                  &                  &                  &   -0.021         &                  &                  &                  &                  &   -0.020         \\
                &                  &                  &                  &                  &  (0.045)         &                  &                  &                  &                  &  (0.046)         \\
            
            Controls  &Y&Y&Y&Y&Y&Y&Y&Y&Y&Y\\  
            State FEs &Y&Y&Y&Y&Y&Y&Y&Y&Y&Y\\
        \hline
        Dependent var. mean &0.358&0.358&0.358&0.330&0.330 &0.471&0.471&0.471&0.443&0.443\\
        
Observations    &      352         &      352         &      352         &      327         &      327         &      244         &      244         &      244         &      219         &      219         \\
R-squared       &    0.341         &    0.343         &    0.343         &    0.349         &    0.337         &    0.295         &    0.295         &    0.295         &    0.318         &    0.308         \\
        \hline\hline
        \end{tabular}
}

	\end{center}
\par
\vspace{0.2cm}
\begin{spacing}{0.75}
{\footnotesize {\scriptsize \textit{Note.} Our sample for this table is restricted to cadets from Slave States. For religion, we use the number of churches per 1,000 population as a proxy for a religious environment. The control variables consist of \textit{Age in 1860}, \textit{Class Rank}, and \textit{Cohort}. Columns (4)–(5) and (9)–(10) have smaller sample sizes because Washington, DC did not have complete voting records prior to 1861. Columns (1)-(5) focus on cadets from Slave States and Columns (6)-(10) on those from Border States. The standard errors reported in the parentheses are derived from bootstrapping with 400 resampling iterations. ***$p<0.01$, **$p<0.05$, *$p<0.1$.}}
\end{spacing}

\end{table}


 
\clearpage

\subsection{Peer Influence by Cohort} \label{A_byCohort}

\begin{figure}[H]
	\begin{center}
	\includegraphics[scale=0.9]{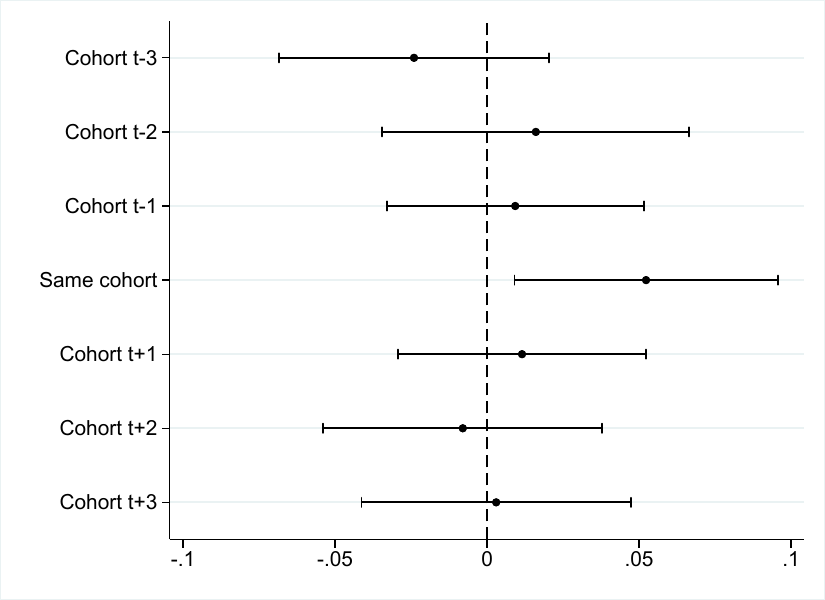}
 \end{center}
        \caption{Peer influence by cohort} \label{fig_bycohort}
        \vspace{0.2cm}
\begin{spacing}{0.75}
{\footnotesize {\scriptsize \textit{Note.} The sample for this analysis consists of the war-participating cadets from Slave States (i.e., with a slave population higher than 1\%). This figure reports the coefficients of peers from the same cohort $t$ and those from three cohorts before and after $t$. The coefficients are derived from a multiple regression that includes all controls from our baseline analysis. The bars represent 90\% confidence intervals.}}
\end{spacing}

\end{figure}

\clearpage
\subsection{Peer Influence and Continuous Military Service Experience}

\begin{table}[h]
		
   \caption{ \centering Peer composition and allegiance choice: Continuous military service experience}\label{Atab_continuous_military}
		
    \begin{center}
			
			\footnotesize 
    \def\sym#1{\ifmmode^{#1}\else\(^{#1}\)\fi}
        \begin{tabular}{l*{6}{c}}
            \hline\hline
            Dependent var
            &\multicolumn{6}{c}{Join the Union: War Participants}
            \\\cmidrule(lr){2-7}
            &\multicolumn{3}{c}{Non-continuous Military Service} 
            &\multicolumn{3}{c}{Continuous Military Service}
            \\\cmidrule(lr){2-4}\cmidrule(lr){5-7}
            &\multicolumn{1}{c}{(1)} 
            &\multicolumn{1}{c}{(2)} 
            &\multicolumn{1}{c}{(3)} 
            &\multicolumn{1}{c}{(4)}
            &\multicolumn{1}{c}{(5)} 
            &\multicolumn{1}{c}{(6)}\\
            \hline
            Share Free-State Peers (sd)    &    0.014         &    0.022         &    0.005         &    0.098\sym{***}&    0.063\sym{**} &    0.062\sym{**} \\
                &  (0.039)         &  (0.035)         &  (0.038)         &  (0.029)         &  (0.026)         &  (0.027)         \\
            Age in 1860        &                  &   -0.006         &    0.013         &                  &   -0.017         &   -0.014         \\
                &                  &  (0.022)         &  (0.024)         &                  &  (0.017)         &  (0.018)         \\
            Class Rank        &                  &   -0.001         &   -0.001         &                  &    0.002\sym{*}  &    0.002\sym{*}  \\
                &                  &  (0.001)         &  (0.001)         &                  &  (0.001)         &  (0.001)         \\
            Slave Pop. Share (sd)&  &      -0.171\sym{***}             &                  &   &      -0.218\sym{***}            \\
                &                  &  (0.030)         &                  &                  &  (0.024)         &                  \\
            Cohort       &                  &   -0.011         &    0.006         &                  &   -0.023         &   -0.020         \\
                &                  &  (0.021)         &  (0.022)         &                  &  (0.016)         &  (0.017)         \\
            
            State FEs &N&N&Y&N&N&Y\\
        \hline
        Dependent var. mean &0.202&0.202&0.202&0.431&0.431&0.431\\
        
Observations    &      119         &      119         &      119         &      269         &      269         &      269         \\
R-squared       &    0.001         &    0.230         &    0.347         &    0.039         &    0.272         &    0.323         \\
        \hline\hline
        \end{tabular}

	\end{center}
\par
\vspace{0.2cm}
\begin{spacing}{0.75}
{\footnotesize {\scriptsize \textit{Note.} The sample for this analysis consists of all war-participating cadets. Columns (1)-(3) use the sample of cadets who remained out of military service for more than one year before the war, while Columns (4)-(6) use the sample of cadets who remained continuously in military service. The standard errors reported in the parentheses are derived from bootstrapping with 400 resampling iterations ***$p<0.01$, **$p<0.05$, *$p<0.1$.}}
\end{spacing}
\end{table}

\clearpage
\subsection{Peer Influence pre and post 1850}

\begin{table}[h]
		
   \caption{ \centering Peer composition and allegiance choice before and after 1850}\label{Atab_1850}

    \begin{center}
			
			\footnotesize 
    \def\sym#1{\ifmmode^{#1}\else\(^{#1}\)\fi}
        \begin{tabular}{l*{4}{c}}
            \hline\hline
            Dependent var
            &\multicolumn{4}{c}{Join the Union: War Participants} 
            \\\cmidrule(lr){2-5}
            &\multicolumn{2}{c}{1820-1849}  
            &\multicolumn{2}{c}{1850-1860}
            \\\cmidrule(lr){2-3}\cmidrule(lr){4-5}
            &\multicolumn{1}{c}{(1)} 
            &\multicolumn{1}{c}{(2)} 
            &\multicolumn{1}{c}{(3)} 
            &\multicolumn{1}{c}{(4)} \\
            \hline
            Share Free-State Peers (sd)   &    0.054\sym{**} &    0.045\sym{*}  &    0.057\sym{*}  &    0.062\sym{*}  \\
                &  (0.026)         &  (0.026)         &  (0.032)         &  (0.034)         \\
            Age in 1860     &   -0.001         &    0.018         &   -0.026         &   -0.030         \\
                &  (0.020)         &  (0.020)         &  (0.022)         &  (0.026)         \\
            Class Rank      &    0.000         &    0.001         &    0.001         &    0.000         \\
                &  (0.001)         &  (0.001)         &  (0.001)         &  (0.001)         \\
            
            Slave Pop. Share (sd)&   -0.224\sym{***}&                  &   -0.206\sym{***}&                  \\
                &  (0.023)         &                  &  (0.032)         &                  \\
            Cohort         &   -0.002         &    0.015         &   -0.013         &   -0.017         \\
                &  (0.020)         &  (0.020)         &  (0.022)         &  (0.026)         \\
            State FEs & N& Y& N& Y\\
            \hline
            Dependent var.mean & 0.389 & 0.389 & 0.318 & 0.318  \\
            Observations    &      234         &      234         &      154         &      154         \\
            R-squared       &    0.236         &    0.301         &    0.231         &    0.332         \\
            \hline\hline
        \end{tabular}

	\end{center}
\par
\vspace{0.2cm}
\begin{spacing}{0.75}
{\footnotesize {\scriptsize \textit{Note.} The sample for this analysis consists of the war-participating cadets from Slave States (i.e., with a slave population higher than 1\%). The standard errors reported in the parentheses are derived from bootstrapping with 400 resampling iterations. ***$p<0.01$, **$p<0.05$, *$p<0.1$.}}
\end{spacing}

\end{table}


\clearpage

\subsection{Controlling for Five-Year Fixed Effects}  
\begin{table}[h]
	
   \caption{ \centering Peer composition and allegiance choice with five-year fixed effects}\label{tab_congressional_rotation}
		
    \begin{center}
			
			\footnotesize     \def\sym#1{\ifmmode^{#1}\else\(^{#1}\)\fi}
        \begin{tabular}{l*{6}{c}}
            \hline\hline
            Dependent var
            &\multicolumn{6}{c}{Join the Union: War Participants}
            \\\cmidrule(lr){2-7}
            &\multicolumn{3}{c}{Slave States} 
            &\multicolumn{3}{c}{Free States}
            \\\cmidrule(lr){2-4}\cmidrule(lr){5-7}
            &\multicolumn{1}{c}{(1)} 
            &\multicolumn{1}{c}{(2)} 
            &\multicolumn{1}{c}{(3)} 
            &\multicolumn{1}{c}{(4)}
            &\multicolumn{1}{c}{(5)} 
            &\multicolumn{1}{c}{(6)}\\
\hline
            Share Free-State Peers (sd) &    0.076\sym{***}&    0.062\sym{***}&    0.058\sym{**} &   -0.008         &   -0.009         &   -0.011         \\
                &  (0.025)         &  (0.022)         &  (0.022)         &  (0.013)         &  (0.012)         &  (0.013)         \\
            Controls &N&Y&Y&N&Y&Y\\
            State FEs &N&N&Y&N&N&Y\\
            Five-Year Interval FEs &Y&Y&Y&Y&Y&Y\\
        \hline
        Dependent var. mean &0.361&0.361&0.361&0.919&0.919&0.919\\
        Observations    &      388         &      388         &      388         &      540         &      540         &      540         \\
        R-squared       &    0.041         &    0.240         &    0.292         &    0.033         &    0.043         &    0.057         \\
        \hline\hline
        \end{tabular}

	\end{center}
\par
\vspace{0.2cm}
\begin{spacing}{0.75}
{\footnotesize {\scriptsize \textit{Note.} This table presents the impact of the fraction of peers from Free States in a cadet's cohort on that cadet's decision to join the Union.  We include fixed effects at five year intervals to allow for a shift in the environment. The control variables consist of \textit{Age in 1860}, \textit{Class Rank}, state level \textit{Slave Population Share (sd)}, and \textit{Cohort}. Columns (1)-(3) focus on cadets from Slave States and Columns (4)-(6) on those from Free States. The standard errors presented in the parentheses are obtained through bootstrapping with 400 resampling iterations. ***$p<0.01$, **$p<0.05$, *$p<0.1$.}}
\end{spacing}

\end{table}
\clearpage

\subsection{Graduate Peers vs. Dropout Peers} \label{A_dropout}


\begin{table}[h]
		
   \caption{ \centering Considering dropout peers}\label{Atab_dropout}

    \begin{center}
			
			\footnotesize 
    \def\sym#1{\ifmmode^{#1}\else\(^{#1}\)\fi}
        \begin{tabular}{l*{4}{c}}
            \hline\hline
            Dependent var
            &\multicolumn{4}{c}{Join the Union: War Participants}
            \\\cmidrule(lr){2-5}
            &\multicolumn{4}{c}{Slave States} 
            \\\cmidrule(lr){2-5}
            &\multicolumn{1}{c}{(1)}
            &\multicolumn{1}{c}{(2)}
            &\multicolumn{1}{c}{(3)}
            &\multicolumn{1}{c}{(4)} \\
            \hline
            Share Free-State Total Peers (sd) &    0.058\sym{**} &    0.055\sym{**} &                  &                  \\
                &  (0.023)         &  (0.023)         &                  &                  \\
            Share Free-State Graduates (sd)    &                  &                  &    0.059\sym{***}&    0.053\sym{**} \\
                &                  &                  &  (0.022)         &  (0.022)         \\
            Share Free-State Dropouts (sd)      &                  &                  &    0.009         &    0.010         \\
                &                  &                  &  (0.021)         &  (0.020)         \\
            Age in 1860    &   -0.011         &   -0.004         &   -0.012         &   -0.005         \\
                &  (0.015)         &  (0.015)         &  (0.015)         &  (0.016)         \\
            Class Rank      &    0.001         &    0.001         &    0.000         &    0.001         \\
                &  (0.001)         &  (0.001)         &  (0.001)         &  (0.001)         \\
            Slave Pop. Share (sd)    &   -0.216\sym{***}&                  &   -0.217\sym{***}&                  \\
                &  (0.019)         &                  &  (0.019)         &                  \\
            Cohort    &   -0.014         &   -0.006         &   -0.015         &   -0.007         \\
                &  (0.015)         &  (0.015)         &  (0.015)         &  (0.015)         \\
            State FEs &N&Y&N&Y\\
\hline
           Dependent var.mean & 0.361 & 0.361 & 0.361 & 0.361  \\
           Observations    &      388         &      388         &      388         &      388         \\
            R-squared       &    0.231         &    0.286         &    0.233         &    0.286         \\
\hline\hline
        \end{tabular}
	
	\end{center}
\par
\vspace{0.2cm}
\begin{spacing}{0.75}
{\footnotesize {\scriptsize \textit{Note.} 
The sample for this analysis consists of the war-participating cadets from Slave States (i.e., with a slave population higher than 1\%). Columns (1)-(2) consider all peers, including those who dropped out. Columns (3)-(4) compare the graduate peers with dropout peers. The standard errors reported in the parentheses are derived from bootstrapping with 400 resampling iterations. ***$p<0.01$, **$p<0.05$, *$p<0.1$.}}
\end{spacing}

\end{table}



\clearpage


\clearpage

\subsection{Multinomial Logit Results}

\begin{table}[h]
		
   \caption{\centering Peer influence and allegiance choice: Multinomial logit results}                \label{Atab_mlogit}

    \begin{center}
			
	\footnotesize \def\sym#1{\ifmmode^{#1}\else\(^{#1}\)\fi}
\begin{tabular}{l*{3}{c}}
\hline\hline
Reference group
     &\multicolumn{3}{c}{Join the Confederacy} \\\cmidrule(lr){2-4}
            &\multicolumn{1}{c}{(1)}         
            &\multicolumn{1}{c}{(2)}         
            &\multicolumn{1}{c}{(3)}       \\
\hline
        Share Free-State Peers (sd): Join the Union           &    0.349\sym{***}&    0.336\sym{***}&    0.358\sym{***}\\
                &  (0.109)         &  (0.109)         &  (0.123)         \\
        Share Free-State Peers (sd): Not in war  &    0.102         &    0.063         &    0.079         \\
                &  (0.094)         &  (0.101)         &  (0.105)         \\
        Controls    &      N         &    Y         &      Y            \\
        State FEs    &      N         &    N         &      Y            \\                
\hline
        Dependent var. mean&1.212&1.212&1.212\\
        Observations    &      670         &      670         &      670         \\
        Pseudo R-squared       &        0.008          &       0.181           &     0.206             \\
\hline\hline
\end{tabular}

	\end{center}
\par
\vspace{0.2cm}
\begin{spacing}{0.75}
{\footnotesize {\scriptsize \textit{Note.}
The sample for this analysis consists of all cadets from Slave States, including those who did not join either army. The results show that peer influence mainly affected whether to join the Union or Confederacy rather than whether to engage in the war. The control variables consist of \textit{Age in 1860}, \textit{Class Rank}, state level \textit{Slave Population Share (sd)}, and \textit{Cohort}. The standard errors reported in the parentheses are derived from bootstrapping with 400 resampling iterations. ***$p<0.01$, **$p<0.05$, *$p<0.1$.}}
\end{spacing}

\end{table}

\clearpage

\subsection{Considering Cash-Crop Employment}

\begin{table}[h]
		
   \caption{ \centering Peer composition and allegiance choice: Cash crops}\label{tab_baseline_cashcrop}
		
    \begin{center}
			
			\footnotesize 
    \def\sym#1{\ifmmode^{#1}\else\(^{#1}\)\fi}
    \resizebox{\textwidth}{!}{
        \begin{tabular}{l*{8}{c}}
            \hline\hline
            Dependent var
            &\multicolumn{8}{c}{Join the Union: War Participants}
            \\\cmidrule(lr){2-9}
            &\multicolumn{4}{c}{Slave States} 
            &\multicolumn{4}{c}{Free States}
            \\\cmidrule(lr){2-5}\cmidrule(lr){6-9}
            &\multicolumn{1}{c}{(1)} 
            &\multicolumn{1}{c}{(2)} 
            &\multicolumn{1}{c}{(3)} 
            &\multicolumn{1}{c}{(4)}
            &\multicolumn{1}{c}{(5)} 
            &\multicolumn{1}{c}{(6)}
            &\multicolumn{1}{c}{(7)} 
            &\multicolumn{1}{c}{(8)}\\
            \hline
            Share Free-State Peers (sd)   &    0.080\sym{***}&    0.058\sym{***}&    0.053\sym{***}&    0.052\sym{**} &   -0.003         &   -0.009         &   -0.011         &   -0.011         \\
                &  (0.023)         &  (0.020)         &  (0.020)         &  (0.020)         &  (0.013)         &  (0.012)         &  (0.012)         &  (0.012)         \\
            Age in 1860     &                  &   -0.013         &   -0.006         &   -0.006         &                  &    0.006         &    0.004         &    0.003         \\
                &                  &  (0.015)         &  (0.015)         &  (0.016)         &                  &  (0.007)         &  (0.007)         &  (0.007)         \\
            Class Rank   &                  &    0.001         &    0.001         &    0.001         &                  &   -0.001\sym{*}  &   -0.001\sym{*}  &   -0.001\sym{*}  \\
                &                  &  (0.001)         &  (0.001)         &  (0.001)         &                  &  (0.000)         &  (0.000)         &  (0.000)         \\
            Slave Pop. Share (sd)&                  &   -0.203\sym{***}&                  &                  &                  &   -0.004         &                  &                  \\
                &                  &  (0.019)         &                  &                  &                  &  (0.014)         &                  &                  \\
            Cash Crop Employment  &                  &   -0.259\sym{***}&   -0.234\sym{***}&   -0.230\sym{***}&                  &   -0.627\sym{***}&   -0.638\sym{***}&   -0.621         \\
                &                  &  (0.056)         &  (0.054)         &  (0.054)         &                  &  (0.187)         &  (0.187)         &  (3.384)         \\
            Cash Crop $\times$ Share Free-State Peers (sd) &                  &                  &                  &    0.027         &                  &                  &                  &    0.033         \\
                &                  &                  &                  &  (0.066)         &                  &                  &                  &  (4.332)         \\
            Cohort      &                  &   -0.014         &   -0.007         &   -0.007         &                  &    0.007         &    0.006         &    0.005         \\
                &                  &  (0.015)         &  (0.015)         &  (0.015)         &                  &  (0.007)         &  (0.007)         &  (0.007)         \\
            State FEs &N&N&Y&Y&N&N&Y&Y\\
        \hline
        Dependent var. mean &0.361&0.361&0.361&0.361&0.919&0.919&0.919&0.919\\
        Observations    &      388         &      388         &      388         &      388         &      540         &      540         &      540         &      540         \\
R-squared       &    0.028         &    0.250         &    0.300         &    0.300         &    0.000         &    0.084         &    0.098         &    0.099         \\
        \hline\hline
        \end{tabular}
}

	\end{center}
\par
\vspace{0.2cm}
\begin{spacing}{0.75}
{\footnotesize {\scriptsize \textit{Note.} This table presents the impact of the fraction of peers from Free States in a cadet's cohort on that cadet's decision to join the Union. Following White (2024), we introduce a variable  \textit{Cash crops}, defined as a dummy equal to 1 if the cadet had a history of employment in cash-crop agriculture. This includes cases where the graduate was recorded as having spent time as a planter (plantation owner) or held any occupation related to the production, processing, sale, or export of cotton, indigo, rice, sugar, or tobacco. Columns (1)-(4) focus on cadets from Slave States and Columns (5)-(8) on those from Free States. The standard errors presented in the parentheses are obtained through bootstrapping with 400 resampling iterations. ***$p<0.01$, **$p<0.05$, *$p<0.1$.}}
\end{spacing}
\end{table}

\clearpage

\subsection{Considering Appointment State Peers}

\begin{table}[h]
	
   \caption{ \centering Peer composition and allegiance choice: Home state and appointed state}\label{tab_appointment_state}
		
    \begin{center}
			
			\footnotesize 
    \def\sym#1{\ifmmode^{#1}\else\(^{#1}\)\fi}
        \begin{tabular}{l*{6}{c}}
            \hline\hline
            Dependent var
            &\multicolumn{6}{c}{Join the Union: War Participants}
            \\\cmidrule(lr){2-7}
            &\multicolumn{3}{c}{Slave States} 
            &\multicolumn{3}{c}{Free States}
            \\\cmidrule(lr){2-4}\cmidrule(lr){5-7}
            &\multicolumn{1}{c}{(1)} 
            &\multicolumn{1}{c}{(2)} 
            &\multicolumn{1}{c}{(3)} 
            &\multicolumn{1}{c}{(4)}
            &\multicolumn{1}{c}{(5)} 
            &\multicolumn{1}{c}{(6)}\\
\hline
            Share Free-State Peers (sd)&    0.066\sym{***}&    0.052\sym{***}&    0.046\sym{**} &   -0.009         &   -0.012         &   -0.014         \\
            Appointed State  &  (0.024)         &  (0.020)         &  (0.020)         &  (0.013)         &  (0.013)         &  (0.013)         \\
            Controls &N&Y&Y&N&Y&Y\\
            State FEs &N&N&Y&N&N&Y\\
        \hline
        Dependent var. mean &0.361&0.361&0.361&0.919&0.919&0.919\\
        Observations    &      388         &      388         &      388         &      540         &      540         &      540         \\
        R-squared       &    0.019         &    0.231         &    0.284         &    0.001         &    0.018         &    0.032         \\
        \hline\hline
        \end{tabular}

	\end{center}
\par
\vspace{0.2cm}
\begin{spacing}{0.75}
{\footnotesize {\scriptsize \textit{Note.} This table presents the impact of the fraction of peers from Free States in a cadet's cohort on that cadet's decision to join the Union. \textit{Share Free-State peers (sd) of appointed state} is the proportion of free companions calculated based on the place of appointment of graduates. The control variables consist of \textit{Age in 1860}, \textit{Class Rank}, state level \textit{Slave Population Share (sd)}, and \textit{Cohort}. Columns (1)-(3) focus on cadets from Slave States and Columns (4)-(6) on those from Free States. The standard errors presented in the parentheses are obtained through bootstrapping with 400 resampling iterations. ***$p<0.01$, **$p<0.05$, *$p<0.1$.}}
\end{spacing}

\end{table}

\clearpage
\subsection{Career Outcomes}
\begin{table}[H]
		
   \caption{ \centering Subsequent outcomes in 1865}\label{tab_careeroutcome}
		
    \begin{center}
			
		\footnotesize	
    \def\sym#1{\ifmmode^{#1}\else\(^{#1}\)\fi}
\begin{tabular}{l*{6}{c}}
    \hline\hline
        &\multicolumn{3}{c}{OLS} 
        &\multicolumn{3}{c}{IV}
        \\\cmidrule(lr){2-4}\cmidrule(lr){5-7}
	&\multicolumn{1}{c}{Rank}  
        &\multicolumn{1}{c}{General}
        &\multicolumn{1}{c}{Died}  
        &\multicolumn{1}{c}{Rank}  
        &\multicolumn{1}{c}{General}
        &\multicolumn{1}{c}{Died}
        \\\cmidrule(lr){2-4}\cmidrule(lr){5-7}
	&\multicolumn{1}{c}{(1)}
        &\multicolumn{1}{c}{(2)}         
        &\multicolumn{1}{c}{(3)}         
        &\multicolumn{1}{c}{(4)}         
        &\multicolumn{1}{c}{(5)}         
        &\multicolumn{1}{c}{(6)}          \\

\hline
A. Slave-State Cadets (\textgreater1\%) \\
\hline
        Joining the Union    &   -1.567\sym{***}&   -0.352\sym{***}&   -0.166\sym{***}&   -1.858\sym{***}&   -0.365\sym{***}&   -0.135\sym{*}  \\
                &  (0.194)         &  (0.054)         &  (0.046)         &  (0.346)         &  (0.099)         &  (0.071)         \\
      
      Controls          &  Y      &  Y        &  Y       &           Y  &  Y        &  Y        \\
        State FEs & Y & Y & Y & Y & Y & Y \\          
\hline
        Dependent var. mean &6.987&0.438&0.206&6.987&0.438&0.206\\
        Observations    &      381         &      381         &      388         &      381         &      381         &      388         \\
        F-statistic &        &          &         &    9.911     &   9.911          &     10.255       \\
        R-squared       &    0.317         &    0.190         &    0.097         &    0.267         &    0.149         &    0.058         \\
\hline
B. Slave-State Cadets (1\%-33\%) \\
\hline
        Joining the Union   &   -1.505\sym{***}&   -0.314\sym{***}&   -0.170\sym{***}&   -1.717\sym{***}&   -0.365\sym{**} &   -0.168\sym{*}  \\
                &  (0.223)         &  (0.061)         &  (0.048)         &  (0.513)         &  (0.147)         &  (0.091)         \\
      
      Controls          &  Y      &  Y         &  Y        &           Y  &  Y         &  Y        \\
        State FEs & Y & Y & Y & Y & Y & Y \\          
\hline
        Dependent var. mean &6.802&0.407&0.184&6.802&0.407&0.184\\
        Observations    &      263         &      263         &      266         &      263         &      263         &      266         \\
        F-statistic &        &          &         &    8.916     &   8.916          &    9.189       \\
        R-squared       &    0.314         &    0.191         &    0.082         &    0.262         &    0.144         &    0.070         \\
        \hline\hline
			\end{tabular}

	\end{center}
\par
\vspace{0.2cm}
\begin{spacing}{0.75}
{\footnotesize {\scriptsize \textit{Note.} The sample for Panel A consists of the war-participating cadets from Slave States (i.e., with a slave population higher than 1\%), and the sample for Panel B excludes those from the High-Slave States who were not significantly influenced by peers. The control variables consist of \textit{Age in 1860}, \textit{Class Rank}, and \textit{Cohort}. Columns (1)-(3) use OLS estimation, while Columns (4)–(6) present IV estimates using birthplace indicators and the share of peers from Free States as instruments for Union affiliation. Columns (1) and (4) feature the military rank as the dependent variable, classified into 11 levels, with General assigned 11 and Third Lieutenant assigned 1. General and Died are binary variables. The rank information on seven cadets is unknown. The standard errors reported in the tables are derived from bootstrapping with 400 resampling iterations. ***$p<0.01$, **$p<0.05$, *$p<0.1$. }}
\end{spacing}

\end{table}

\clearpage

\subsection{Correlation between Academic Rank and Military Rank in 1865} \label{A_academic}
\begin{table}[h]
		
   \caption{\centering Academic rank and military rank in 1865 (The Confederacy vs. The Union)}  \label{Atab_academic}

    \begin{center}
			
			\footnotesize \def\sym#1{\ifmmode^{#1}\else\(^{#1}\)\fi}
		\begin{tabular}{l*{6}{c}}
			\hline\hline
                Dependent var
                &\multicolumn{6}{c}{Rank in 1865: War Participants}    
                \\\cmidrule(lr){2-7}
                &\multicolumn{3}{c}{The Confederacy}                     
                &\multicolumn{3}{c}{The Union}
                \\\cmidrule(lr){2-4}\cmidrule(lr){5-7}
                &\multicolumn{1}{c}{(1)}         
                &\multicolumn{1}{c}{(2)}         
                &\multicolumn{1}{c}{(3)}         
                &\multicolumn{1}{c}{(4)}         
                &\multicolumn{1}{c}{(5)}         
                &\multicolumn{1}{c}{(6)} \\
\hline
        Class Rank       &    0.000         &   -0.001         &   -0.001         &    0.007\sym{**} &    0.007\sym{***}&    0.006\sym{**} \\
                &  (0.003)         &  (0.003)         &  (0.003)         &  (0.003)         &  (0.003)         &  (0.003)         \\
        Age in 1860     &                  &   -0.008         &   -0.027         &                  &   -0.031         &   -0.037         \\
                &                  &  (0.052)         &  (0.059)         &                  &  (0.042)         &  (0.046)         \\
        
        Slave Pop. Share (sd)&                  &    0.157\sym{*}  &                  &                  &    0.033         &                  \\
                &                  &  (0.089)         &                  &                  &  (0.064)         &                  \\
        Cohort    &                  &   -0.053         &   -0.074         &                  &   -0.085\sym{**} &   -0.092\sym{**} \\
                &                  &  (0.050)         &  (0.057)         &                  &  (0.040)         &  (0.044)         \\
        State FEs    &      N   &    N        &      Y      &    N   &    N       &      Y      \\
\hline
        Dependent var.mean    &      7.516        &      7.516         &      7.516         &           5.851         &      5.851         &      5.851                \\
        Observations    &      281         &      281         &      281         &      636         &      636         &      636         \\
        R-squared       &    0.000         &    0.086         &    0.163         &    0.010         &    0.111         &    0.129         \\
\hline\hline
		\end{tabular}

	\end{center}
\par
\vspace{0.2cm}
\begin{spacing}{0.75}
{\footnotesize {\scriptsize \textit{Note.} 
The sample for this analysis consists of individuals who served in the Confederate Army (Columns (1)-(3)) and those who served in the Union Army (Columns (4)-(6)). This table shows that the association between academic rank and military rank is significant in the Union but not in the Confederacy. The standard errors reported in the parentheses are derived from bootstrapping with 400 resampling iterations. ***$p<0.01$, **$p<0.05$, *$p<0.1$.}}
\end{spacing}

\end{table}



\end{document}